\documentclass[reprint,superscriptaddress,amsmath,amssymb,aps,longbibliography]{revtex4-2}

\usepackage{amsmath}
\usepackage{amssymb}
\usepackage{amsthm}
\usepackage{amsfonts}
\usepackage[caption=false]{subfig}
\usepackage[colorlinks, citecolor=blue]{hyperref}
\usepackage{tikz}
\usepackage{relsize}
\usepackage{color,soul}
\usepackage[utf8]{inputenc}
\usepackage{capt-of}
\usepackage{mathtools}
\usepackage{float}
\usepackage[section]{placeins}
\usepackage{listings}
\usepackage{svg}
\usepackage{siunitx}
\usepackage{hhline}
\usepackage[english]{babel}

\usepackage[T1]{fontenc}  
\usetikzlibrary{decorations.pathreplacing}
\usepackage{braket}

\theoremstyle{definition}

\theoremstyle{definition}

\theoremstyle{definition}

\newcommand{\eq}[1]{\hyperref[eq:#1]{Equation~\ref*{eq:#1}}}
\renewcommand{\sec}[1]{\hyperref[sec:#1]{Section~\ref*{sec:#1}}}
\newcommand{\fig}[2][]{\hyperref[fig:#2]{Fig.~\ref*{fig:#2}\textbf{#1}}}
\newcommand{\tbl}[1]{\hyperref[tbl:#1]{Table~\ref*{tbl:#1}}}
\newcommand{\theoremref}[1]{\hyperref[theorem:#1]{Theorem~\ref*{theorem:#1}}}
\newcommand{\definitionref}[1]{\hyperref[definition:#1]{Definition~\ref*{definition:#1}}}

\begin{document}

\title{Reinforcement learning control of quantum error correction}

\author{Google Quantum AI and Google DeepMind}

\begin{abstract}
    Quantum error correction (QEC) is the primary strategy for protecting a quantum computer from the environment \cite{shor1995scheme,nielsen2010quantum}. Its prerequisite is that errors must remain sufficiently rare, which requires perpetually adapting the computer’s control parameters to the drifting environment conditions. The current solution to this problem is to terminate the entire quantum computation for recalibration, but it is incompatible with the long runtimes of future quantum algorithms \cite{reiher2017elucidating,gidney2025factor}. We address this challenge by unifying calibration with computation. We grant the QEC process \cite{ryan2021realization,krinner2022realizing,sivak2023real,acharya2024quantum, bluvstein2024logical,bluvstein2025architectural,lacroix2025scaling} a dual role: its error detection events are not only used to correct the logical quantum state, but are also repurposed as a learning signal, teaching a reinforcement learning (RL) agent \cite{silver2017mastering,mnih2015human,levine2016end, shalev2016safe,ouyang2022training} to continuously steer the control parameters and stabilize the quantum system during computation. We experimentally demonstrate this framework on a Willow superconducting processor, improving the logical stability of the surface code 3.5-fold against injected drift. By synthesizing our full suite of technological advances, we achieve record performance of the surface and color codes, with average logical error per cycle of $7.72(9)\times10^{-4}$ and $8.19(14)\times10^{-3}$ respectively. Numerical simulations of large codes with tens of thousands of control parameters confirm the scalability of our RL framework, revealing an optimization speed that is independent of system size. This work thus enables a new paradigm: a quantum computer that learns from its errors and never stops computing.
\end{abstract}

\maketitle

\setlength{\textfloatsep}{10pt}
\setlength{\floatsep}{2pt}
\setlength{\intextsep}{2pt}

Quantum computers are fundamentally analog machines, which makes them extremely fragile compared to the digital classical computers \cite{landauer1995quantum}. 
QEC mitigates this vulnerability by effectively digitizing the errors:
through repetitive error detection embedded in QEC protocols, spurious analog evolution branches into a sequence of ``error'' or ``no error'' events.
These binary error detection signals can then be decoded to correct the logical quantum state, providing a mechanism for reaching low logical error rates required for practical applications.

However, the digitization of errors alone is insufficient, and the practical success of QEC is critically dependent on the analog control of the constituent qubits: QEC is effective only if the physical gate error rate is kept significantly below a certain threshold, around $10^{-3}-10^{-2}$ \cite{stephens2014fault}. The process of precisely tuning the system's control parameters to meet this prerequisite is known as calibration \cite{khaneja2005optimal,kelly2018physical}. While traditional calibration techniques have been able to achieve the necessary precision \cite{acharya2024quantum,bluvstein2025architectural}, the analog nature of the control makes the system performance susceptible to drift. 
Consequently, the challenge expands from merely reaching deep below the QEC threshold to continuously maintaining this performance amidst the non-stationarity.

\begin{figure}
    \centering
    \includegraphics[width=\columnwidth]{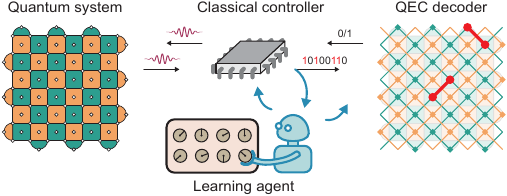}
    \caption{{\bf Overview of RL control.} 
    Quantum computation on logically encoded states is physically realized via analog control signals. The quantum errors are digitized by the QEC process, and error detection events are used by the decoder to infer logical corrections. In our control framework, they are also repurposed as a learning signal teaching the RL agent  to continuously steer the physical control parameters and stabilize the quantum system during computation.}
    \label{fig1}
\end{figure}

The solution to the drift problem used in prior experiments \cite{ryan2021realization,krinner2022realizing,sivak2023real,acharya2024quantum, bluvstein2024logical,bluvstein2025architectural} is to terminate the entire QEC process for intermittent system recalibration. However, this complete decoupling of computation and calibration represents a fundamental bottleneck for useful future algorithms that require continuous execution times on the order of days or even months \cite{reiher2017elucidating,gidney2025factor}. Although theoretical proposals based on logical swaps or code deformation \cite{fang2024caliscalpel} have been put forward to reconcile this dichotomy, they entail a large overhead in the circuit runtime, footprint, and operational complexity. Yet, the very process of error detection embedded in the QEC algorithm offers a more direct solution, since errors from imperfect calibrations produce detectable syndromes just like all other errors. A pioneering attempt in Ref.~\cite{kelly2016scalable} to leverage this by engineering a direct feedback loop from error detection to physical control relied on a heuristic approach that is difficult to scale. In this work, we propose a different paradigm: repurposing error detection events as a learning signal for an artificial intelligence agent, see Fig.~\ref{fig1}.

Our framework is based on reinforcement learning, an approach that demonstrated remarkable success in automating solutions to complex control problems across diverse fields. It has powered systems that achieve superhuman performance in intricate games \cite{silver2017mastering,mnih2015human}, enabled significant advances in robotics \cite{levine2016end, shalev2016safe}, and has recently become integral to refining the behavior of large language models \cite{ouyang2022training}. In the field of quantum control, RL was theoretically explored for various problems \cite{fosel2018reinforcement,bukov2018reinforcement,sivak2022model,metz2023self,bukov2026reinforcement} and applied in experiments to improve the performance of isolated gates \cite{ding2023high, baum2021experimental, reuer2023realizing} and bosonic codes \cite{sivak2023real}. 
Building on these foundations, here we demonstrate the effectiveness of RL for the system-level challenge of calibrating and steering a large error-corrected quantum processor.

We perform our experiments on the distance-5 and 7 surface codes and distance-5 color code, focusing on a quantum memory algorithm where the logical state is being preserved over time via repetitive application of QEC cycles. Our RL agent manages over a thousand control parameters which specify how an abstract QEC circuit is translated into analog waveforms that control the quantum system. The agent successfully steers the system against injected drift, improving the logical error rate (LER) stability $2.4$-fold, a figure that increases to $3.5$-fold with complementary decoder steering. Moreover, RL fine-tuning of an already well-calibrated  processor achieves an additional $20\%$ suppression of the LER, pushing performance beyond the limits of traditional physics-based calibration and human-expert tuning. The agent is able to reach high performance even starting from randomized initial control parameters, suggesting the potential to augment or replace elements of the traditional calibration stack.

Having demonstrated RL control of today's state-of-the-art QEC processor, we further showcase the future potential of this framework using numerical simulations. They establish the compatibility of RL control with uninterrupted quantum computation and its scalability to distance-15 surface code with tens of thousands of control parameters. Our approach is fundamentally general, requiring only error detection signals and tunable controls. Although demonstrated here using a planar QEC code implemented with superconducting circuits, it is directly applicable to any physical qubit modality and QEC architecture, including those with spatially nonlocal connectivity \cite{Bravyi2024}. Our findings thus firmly establish RL as a promising path towards automating the control of large-scale error-corrected quantum systems.

\section*{Learning from errors}

\begin{figure}
    \centering
    \includegraphics[width=\columnwidth]{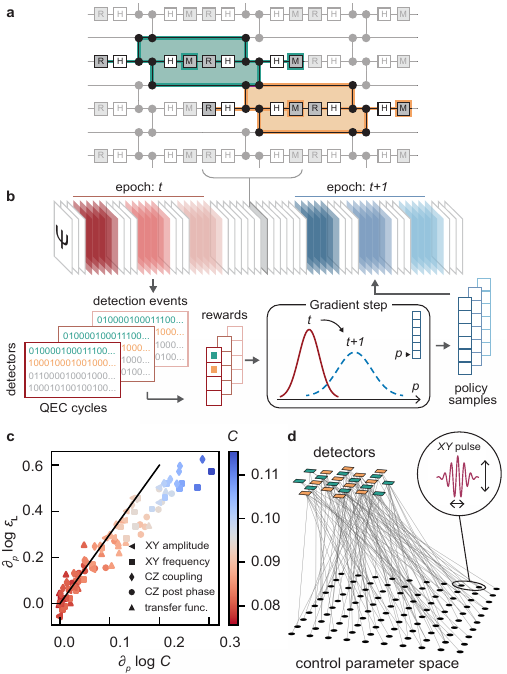}
    \caption{{\bf Learning from errors.} 
    {\bf (a)} A small space-time chunk of the QEC circuit for the repetition code, highlighting two overlapping detecting regions.
    {\bf (b)} One iteration of the learning process. In each epoch, a batch of control policy candidates is sampled from the policy distribution. A certain number of QEC cycles is executed with each policy candidate (shades of red and blue). The acquired QEC data is used to compute rewards by estimating error detection rates for each detector. This information, indicating the relative performance of each policy candidate, is converted by the learning algorithm into a small gradient step for the policy distribution. Then, a new batch of policy candidates is sampled and the process repeats.  
    {\bf (c)} Finite-difference partial derivatives experimentally evaluated along random directions in the high-dimensional control space confirm the linear relation between the gradient of the true and surrogate objectives, with proportionality coefficient $(d+1)/2$ (black line). 
    {\bf (d)} The surrogate objective $C$ allows us to effectively utilize the sparse dependence of error detection rates on the system control parameters, represented here as a factor graph. Each detector node is connected to the learnable control parameters of the gates within its detecting region. In our distance-5 surface code experiment, on average each detector node is connected to $302$ parameter nodes, and each parameter node to $18$ detector nodes. 
    }
    \label{fig2}
\end{figure}

The main guiding principle in the construction of fault-tolerant QEC circuits is to ensure that errors leave detectable signatures in the measurement outcomes of the code stabilizers. These signatures are most conveniently represented using the notion of detectors \cite{gidney2021stim}, defined as sets of measurements that have deterministic total parity in the absence of errors. 
A detection event, corresponding to flipped parity, signals occurrence of errors in a certain space-time region of the circuit, called the detecting region \cite{mcewen2023relaxing}, see Fig.~\ref{fig2}(a). Detecting regions are usually sensitive to multiple kinds of errors, and a decoder is needed to infer the likely error patterns from any given pattern of detection events \cite{higgott2023sparse, beni2025tesseract}. After the decoder's correction of the logical state, the remaining error is quantified by the LER -- the principal measure of quality of the QEC process.

To execute QEC, the parameters of the quantum gates are stored in the memory of a classical controller, see Fig.~\ref{fig1}. These parameters are used by the pulse compiler to convert an abstract QEC logical circuit to its physical implementation. In our learning-based approach, this process is modified: we deliberately apply small, simultaneous perturbations to all control parameters during the computation to explore the control space. These perturbations translate into subtle changes in the statistics of error detection events. The goal of the learning algorithm is to disentangle cause and effect, correlating specific perturbations with changes in performance to continuously calibrate the system.

The performance measure of interest in this process is the logical error rate, $\varepsilon_{\rm L}$. However, using it directly as an optimization objective faces fundamental obstacles to scalability, see Methods. Instead, we construct a surrogate objective $C$, which is an efficiently-computable local proxy for LER that enables high-dimensional optimization. 
Our surrogate objective is defined as the average rate of error detection events, $C=\hat{\mathbb{E}}[D]$, where $D$ denotes all detectors in the circuit, and the empirical expectation value is computed over a sufficient number of QEC cycles. 
This surrogate objective function is inspired by a simple scaling model of the surface code $\varepsilon_{\rm L}\propto\Lambda^{-d/2}$, where $d$ is the code distance and $\Lambda=\varepsilon_{\rm th}/\varepsilon$ quantifies the suppression of the physical error rate $\varepsilon$ relative to the QEC threshold $\varepsilon_{\rm th}$. Since $C\propto\varepsilon$, the gradients of LER and of the surrogate objective in this model are simply related as $\nabla \log \varepsilon_{\rm L} = (d+1)/2\cdot\nabla\log C$. To confirm that this relation is approximately satisfied in a real experimental setting, we sample Gaussian perturbations in the control parameter space and evaluate finite-difference partial derivatives, see Fig.~\ref{fig2}(c). We observe good empirical agreement in the regime of small perturbations, which is sufficient for our RL approach that relies on Monte Carlo gradient estimation.

To optimize $C$, we employ a multi-objective policy-gradient reinforcement learning algorithm, see Methods. 
Key to its performance and scalability is an observation that components of $C$ have sparse dependence on the system control parameters owing to the locality of detecting regions in the QEC circuit, see Fig.~\ref{fig2}(d).
Our algorithm maintains a parameterized probability distribution over all system control parameters \cite{sehnke2010parameter}. Each sample from this distribution, the so-called control policy, is a candidate solution to the optimization problem. As illustrated in Fig.~\ref{fig2}(b), the learning proceeds iteratively, by repeating the same sequence of steps in every learning epoch. First, a batch of policy candidates is sampled and ranked according to $C$: policies that lead to a lower rate of error detection events receive a higher reward. In practice, the error detection rate is estimated by acquiring data from a finite number of QEC cycles executed with a given control policy, and then switching to another policy. 

Next, the reward information, reflecting the relative performance of all policy candidates, is used by the learning algorithm to compute the gradient of the policy distribution, pushing it towards a more optimal region in the control space. 
Due to the limited data rate that currently hampers our ability to learn complex policy distributions such as neural networks, we resort to a simple factorized multivariate Gaussian distribution.
Its mean $\mu(t)$ represents the best guess of the optimal control policy during the learning epoch $t$, and the diagonal covariance $\sigma(t)^2$ controls the degree of exploration of the control space. It evolves during the learning process, typically shrinking over time in order to finely localize the solution \cite{sivak2022model}. Eventually, the distribution converges to a local optimum of the optimization objective, in which detection event probabilities are minimized, indicating that occurrence of errors is suppressed to its minimum. In a non-stationary setting, the stochastic gradients steer the control parameters to follow the system drift. In this case, $\mu(t)$ learns to track the optimal policy over time, while $\sigma(t)^2$ maintains finite spread to never cease exploring.

\section*{RL fine-tuning of control policy}

To demonstrate the practical utility of our RL framework, we first show that it can push system performance beyond the limits achievable by the traditional calibration stack and human-expert tuning. 

The central idea in the traditional approach to gate calibration is to orthogonalize the space of control parameters, using an accurate physical model of the system, and then construct dedicated ``calibration experiments'' that target a few parameters at a time. For example, a control parameter corresponding to the frequency of a microwave pulse for XY-rotation of a qubit is calibrated via spectroscopy, and then the amplitude of this pulse is calibrated via Rabi oscillations. Control parameters can also be jointly fine-tuned in calibration experiments that measure an aggregate performance metric of the gate, such as randomized or cross-entropy benchmarking \cite{knill2008randomized,arute2019quantum}.
Recently, this approach has culminated in a framework for automated calibration of large quantum systems based on the traversal of a directed acyclic graph \cite{kelly2018physical}, which encodes the order of calibrations of various control parameters, distilled from decades of research in the field of quantum control \cite{koch2022quantum}. It has been successfully applied in numerous experiments \cite{arute2019quantum,wu2021strong,acharya2024quantum,lacroix2025scaling} reaching the scale of $O(10^2)$ qubits.

\begin{figure}
    \centering
    \includegraphics[width=\columnwidth]{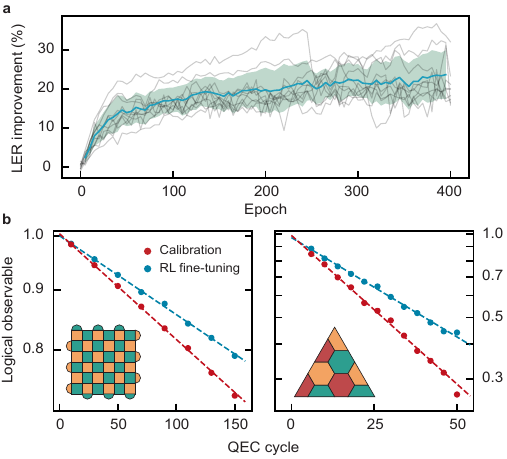}
    \caption{{\bf RL fine-tuning of QEC performance.} 
    {\bf (a)} Systematic improvement of LER from RL fine-tuning applied after exhaustive conventional calibration process, with five independent runs for surface code and color code each (grey), mean performance (teal) and one-sigma deviation (shaded region).
    {\bf (b)} Decay of the logical observable in a quantum memory experiment, averaged over $X$ and $Z$ logical bases, featuring enhanced performance from RL fine-tuning. The distance-7 surface code was decoded with AlphaQubit2, and distance-5 color code with Tesseract.
    }
    \label{fig3}
\end{figure}

Using this traditional approach and extensive human-expert tuning, we calibrate the Willow quantum processor for QEC of the surface code and color code, as in similar prior experiments \cite{google2023suppressing,acharya2024quantum,eickbusch2025demonstratingdynamicsurfacecodes,lacroix2025scaling}.
Subsequent RL fine-tuning of the system consistently yields about $20\%$ additional suppression of the LER, as shown in Fig.~\ref{fig3}(a) for multiple runs on the distance-5 codes of both types. 
We achieve record QEC performance demonstrated across any physical qubit modality: our distance-7 surface code reaches the LER of $\varepsilon_L=7.72(9)\times10^{-4}$ with the AlphaQubit2 neural network decoder \cite{senior2025scalable} (averaged over $X$ and $Z$ logical bases), while the distance-5 color code reaches $\varepsilon_L=8.19(14)\times10^{-3}$ with the Tesseract most-likely-error decoder \cite{beni2025tesseract}, see Fig.~\ref{fig3}(b). 
Our RL fine-tuning technique was also applied in the recent magic state cultivation experiment \cite{rosenfeld2025magic} which combines the elements of both surface code and color code QEC, where it provided an order of magnitude improvement in the cultivation error and the postselection rate.
Beyond fine-tuning, in Supp.~Mat.~IV we show that RL successfully recovers the performance even after we artificially randomize the control policy to fully scramble the logical observable. 

Generally, as gate error rates are pushed deeper below the QEC threshold, they will become limited by a growing number of low-probability error channels. These can be related to the unavoidable simplifying assumptions and approximations in the system models used by the standard calibration approach, or to yet unknown device physics. We anticipate that in the deep below-threshold regime, achieving and maintaining optimal system performance through metrology and targeted calibrations alone will become infeasible, and holistic model-free in-context fine-tuning of the QEC system will become requisite. Our results indicate that RL from error detection events could fill this critical role.

\section*{RL steering of control policy}

\begin{figure}
    \centering
    \includegraphics[width=\columnwidth]{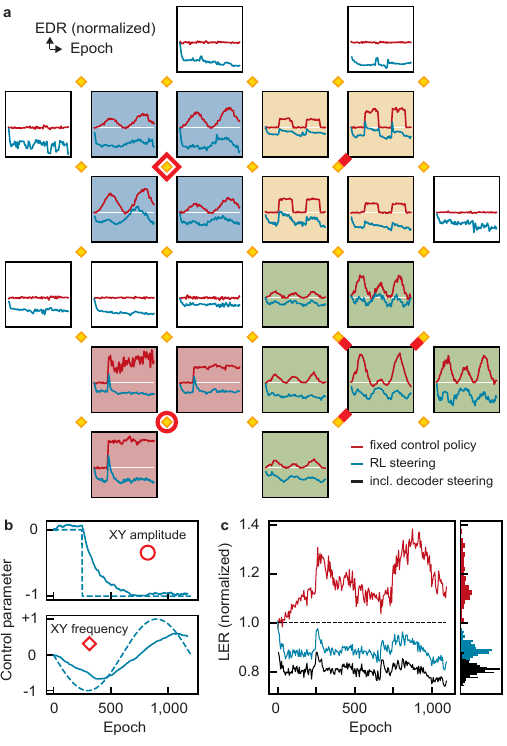}
    \caption{{\bf Demonstration of RL steering.} 
    {\bf (a)} The data qubits (gold diamonds) and measure qubits (panels with data) are arranged in the layout of a distance-5 surface code. We inject artificial drift of various temporal profiles on the gates indicated with red shapes (circle, diamond, bars) and observe elevated error detection signals where expected (colored background). The detection rate associated with each measure qubit is normalized for visualization to remove the effect of the natural system drift. Performance of the fixed control policy (maroon) degrades over time due to the injected drift, while RL steering (blue) stabilizes and maintains the error detection rate (EDR) below its initial level (white lines).
    {\bf (b)} Time-dependence of the injected drift in system control parameters (dashed) and RL steering (solid). Exponential fit of RL response to step-like drift in XY pulse amplitude yields the characteristic learning time of $130$ epochs. 
    {\bf (c)} Periodic evaluation of the logical performance indicates that RL steering of the system significantly suppresses and stabilizes the LER, see main text. Additionally incorporating the decoder steering (black) further improves these results.
    }
    \label{fig4}
\end{figure}

Next, we demonstrate that RL is not only able to optimize the system performance, but also steer the control policy in the presence of drift. This is achieved via the entropy regularization technique \cite{haarnoja2018soft} which ensures that the policy distribution never becomes deterministic, enabling the agent to continuously explore and adapt to changes in a nonstationary environment. To systematically study this effect, we inject artificial drift on several control parameters simultaneously, shown with red symbols in Fig.~\ref{fig4}(a). We choose different temporal drift profiles (step-like, sinusoidal, and stroboscopic), different control parameters (CZ coupling strength, XY pulse amplitude, and frequency), and different locations of drifting gates on the qubit grid. 

Following the already-described methodology, we start by calibrating the control policy at $t=0$. The performance of this policy (maroon) degrades over time due to the injected drift, since the fixed values of the control parameters become ``outdated'', causing additional errors. These control errors lead to an elevated detection event signal, which appears exactly in those detectors where it is expected based on the constructed factor graph, see tiles highlighted with colored background in Fig.~\ref{fig4}(a). 
In contrast to a fixed control policy, RL steering (blue) maintains a significantly suppressed rate of error detection events that consistently remains below the initial level (white line), except for brief moments when the drift is too fast.
In Fig.~\ref{fig4}(b), we show the evolution of the learned control parameters. The recovery from a step-like drift in the XY pulse amplitude (red circle) allows us to estimate the response time of the steering process of about $130$ epochs. This also sets the time scale for the policy lag in the case of slow continuous drift, as in XY pulse frequency (red diamond).

To confirm that suppression of detection events is not due to hindered detection capability but is in fact due to suppressed errors, we evaluate the logical performance in Fig.~\ref{fig4}(c).
Compared to the fixed policy, we find on average a $24\%$ reduction of LER and a $2.4$x improvement of its stability (quantified here by the standard deviation of the LER distribution). These figures of merit further improve to $31\%$ and $3.5$x respectively, by additionally steering the decoder parameters. Decoder steering is achieved within the same RL framework by reweighting the matching graph as described in Ref.~\cite{sivak2024optimization}. While steering of the classical controller is done via the surrogate objective $C$ that relies exclusively on the error detection probabilities, our decoder steering process relies on LER estimation, which is not straightforwardly scalable to the real-time setting. However, alternative approaches to decoder steering have been proposed \cite{google2021exponential,wang2023dgr,hockings2025improving} that, in principle, do not suffer from this limitation.

We also analyzed the RL performance under natural system drift, which arises from sources ranging from material defects near the quantum system \cite{klimov2018fluctuations} to temperature fluctuations in the classical control instruments and, unlike our prior demonstration, rarely has a simple time dependence. Fourier analysis of multiple experimental RL runs from Fig.~\ref{fig3}(a) reveals that the effect of steering can be understood as a filter that provides about $4\;\rm dB$ of suppression of low-frequency LER fluctuations originating from these natural sources, see Supp.~Mat.~III.

We have thus far demonstrated that the mean $\mu(t)$ of the Gaussian policy distribution learns to track the optimal policy over time in the presence of drift. As a result, the learned policy $\mu(t)$ has superior performance compared to the fixed policy, with lower EDR and LER. However, since the learning process requires exploration of the parameter space, the algorithm inevitably samples policy candidates whose performance is worse than that of the $\mu(t)$ policy. This ``exploration noise'' is irrelevant in our experimental setting where RL steering relies on repetitive executions of a short logical algorithm and the quantum state is independently re-prepared in every shot. However, in the future this steering must be done in real time during the single-shot execution of a long logical algorithm. In that case, the exploration noise, although necessary for learning, is going to be detrimental to the logical algorithm's performance. Generally, balancing the exploration of parameter space and exploitation of the learned policy $\mu(t)$ is a central challenge in many applications of RL \cite{sutton2018reinforcement}. In our case, the favorable resolution of the exploration-exploitation tradeoff would mean that aggregate performance of all sampled policy candidates, most of which are worse than $\mu(t)$, is still better than the performance without RL steering.

To study this tradeoff, we conducted numerical simulations of real-time steering of the distance-3 surface code subject to sinusoidal parameter drift at different frequencies, see Fig.~\ref{fig5}(a) and Supp.~Mat.~VI~A. We count the total number of error detection events generated in a $1.8\times 10^9$-cycle window of QEC, and normalize it so that level 1 corresponds to performance of the optimal policy (known in the simulation), and level 0 corresponds to the performance of a fixed policy (optimal at $t=0$). We control the exploration-exploitation tradeoff by changing the amount of entropy regularization \cite{haarnoja2018soft} in our RL algorithm. 
Our findings indicate that there is a critical drift frequency, about $1/150$ epochs, below which the system becomes real-time steerable: the performance with exploration noise is better than the performance of a fixed policy. This critical frequency is consistent with the response time of the learning algorithm observed in experiment, see Fig.~\ref{fig4}(b). When drift is too fast, the real-time steering is not able to keep up -- such drift must be mitigated at the hardware level. In particular, correlated drift caused by rare high-energy particle impacts in superconducting devices \cite{google2023suppressing, acharya2024quantum} typically equilibrates on a much shorter time scale and is not real-time steerable in our current implementation. In contrast, at low drift frequencies, the exploration and exploitation can be successfully balanced to closely approach the performance of the optimal policy at all times. 

Thus, our simulations establish that RL is able to effectively utilize the information concealed in the error detection events to calibrate the system while continuing the logical computation. This ability offers a significant advantage over approaches based on the synthesis of traditional calibration and code deformation \cite{fang2024caliscalpel}, since RL steering does not introduce any resource overhead.

\section*{Scaling and outlook}

\begin{figure}
    \centering
    \includegraphics[width=\columnwidth]{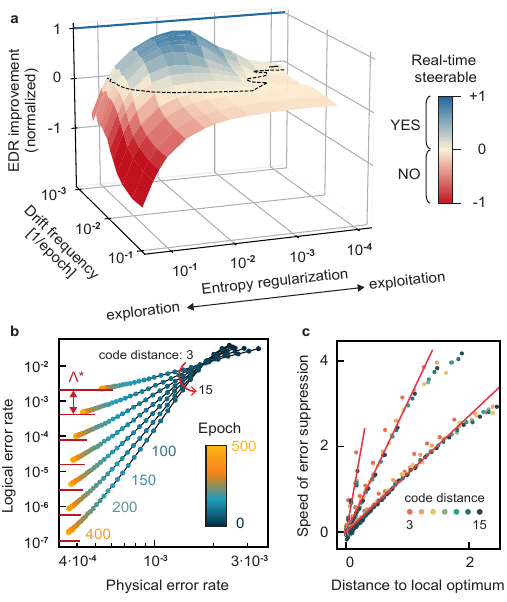}
    \caption{{\bf Real-time steering and scaling simulations.} 
    {\bf (a)} Normalized improvement (color) of error detection rate in the real-time steering simulation of the distance-3 surface code subject to sinusoidal drift at different frequencies. Level $1$ indicates the performance of optimal policy. Isoline at level $0$ demarcates the boundary beyond which real-time steering results in better performance than a fixed policy, approaching the performance of optimal policy in the regime of slow drift.
    {\bf (b)} Simulation of scalability of RL control of large surface codes. The algorithm learns the parameters of single-qubit and CZ gates, with $30$ control parameters per gate, amounting to almost $40,000$ control parameters in total for the distance-15 code. During the learning process, the logical error rate reduces over time (color) until it reaches the floor (red bars) set by the irreducible physical error rates and characterized by the optimal error suppression factor $\Lambda^*$. 
    {\bf (c)} Point-estimates of $\Lambda$ at every code distance and learning epoch from (b) confirm that the speed, $\partial_t \Lambda / \Lambda^* \times 10^2$, at which the error suppression factor approaches the local optimum, is proportional to the distance from the optimum, $1-\Lambda/\Lambda^*$. The convergence rate $\gamma$ (see main text) is independent of the system size but depends on the number of control parameters per gate, with three beams corresponding to 1, 10, and 30 parameters, and the linear fits (red) indicating the convergence rates.
    }
    \label{fig5}
\end{figure}

Finally, we demonstrate the scalability of RL control with simulations extending to a distance-15 surface code with approximately $40,000$ control parameters, see Fig.~\ref{fig5}(b). We extract point-estimates of the instantaneous error suppression factor $\Lambda$ at each code distance and learning epoch, verifying in Fig.~\ref{fig5}(c) that its convergence rate $\partial_t \Lambda/\Lambda^*$ is proportional to the distance from the optimal $\Lambda^*$, set in our simulation by the irreducible control-independent physical error rates. This empirical result leads to the exponential convergence law, $1-\Lambda/\Lambda^*\propto e^{-\gamma\,t}$, characteristic of gradient-descent based algorithms in the vicinity of local optimum \cite{boyd2004convex}, with the convergence rate $\gamma$ that depends on properties such as the number of gates per detecting region and the number of control parameters per gate. Crucially, this rate is independent of the system size, a direct consequence of our algorithm's ability to effectively utilize the sparsity of the factor graph representation of the control problem. 

Multiple possible extensions of our RL control framework leave ample room to further enhance its capabilities. For example, instead of relying on a simple Gaussian policy distribution, it could be advantageous to employ deep neural networks \cite{mnih2015human} and condition the policy distribution on the observations from the environment, e.g. on some suitable statistics of the error detection events. Learning the system model in addition to the policy \cite{hafner2019dream} can lead to better sample efficiency and further improve the quality of solutions by discovering unexpected correlations in the data. It is conceivable that in the future, with sufficient enhancement of our learning framework, a quantum processor could be calibrated for QEC {\it ab initio} fully by RL, with no reliance on the traditional calibration paradigm or human experts. 

In conclusion, by empowering the quantum computer to learn from its errors, we unlock a scalable pathway to optimize performance in real time, replacing disruptive calibration routines with uninterrupted computation. Our work suggests that the path to fault tolerance will be built not only on better hardware, but on more intelligent control.

\section*{Contributions}
V.S. conceived and led the project, with significant contribution from A.M; 
V.S. developed the RL control framework;
A.M. and R.G.C. calibrated and expert-tuned the processor;
V.S. and A.M. designed and conducted the experiments and wrote the manuscript with input from all authors;
M.B. and M.N. contributed software improvements;
N.S., L.A.B. and DeepMind team contributed high-accuracy decoding;
A.E. performed device simulations;
P.K. supervised the project. 
All authors contributed to building and maintaining the hardware, software, cryogenics, and electronics infrastructure.

\section*{Corresponding authors}
Correspondence and requests for materials should be addressed to Volodymyr Sivak (vladsivak@google.com) and Alexis Morvan (amorvan@google.com).

\section*{Competing Interests}
Author-affiliated entities have filed US and international patent applications related to RL calibration of the QEC decoder and quantum computer control parameters, including US18/680,288 and US19/345,222.

\section*{Data availability}
The data supporting our findings are available on Zenodo at \href{https://doi.org/10.5281/zenodo.17566521}{https://doi.org/10.5281/zenodo.17566521}.

\section*{Code availability}
The custom code used in this study is the proprietary property of Google LLC and cannot be made publicly available. Detailed mathematical description of the RL algorithm is provided in Supp. Mat. VIII  to allow independent replication.

\section*{Methods}\label{methods}

\subsection*{Objective function}\label{methods:rl_algo}

Using LER as an optimization objective faces fundamental obstacles to scalability, particularly in multi-qubit stabilizer codes such as the surface code, for the following reasons: 
(i) The LER is suppressed with the code distance $d$ as $\varepsilon_{\rm L}\propto\Lambda^{-d/2}$, where the error suppression factor $\Lambda>2$ has already been demonstrated in experiments with superconducting circuits \cite{acharya2024quantum} and neutral atoms \cite{bluvstein2025architectural}. Thus, accurately resolving the LER would require an exponentially large number of QEC cycles.
(ii) The optimization involves a vast number of parameters -- already more than two thousand in our $d=7$ experiment, and scaling as $O(d^2)$ -- which renders global optimization from a single scalar metric impractical.
(iii) The LER is unsuitable for real-time calibration and steering during a quantum computation, as the logical state is generally unknown.
While direct LER optimization has proven effective in specific contexts, such as for small bosonic codes \cite{sivak2023real}, these scaling challenges necessitate a different approach for large multi-qubit codes.

The surrogate objective $C$, defined in the main text, immediately alleviates the limitation (iii). It also alleviates (i), since resolving $C$ to a constant relative accuracy requires a number of QEC cycles that scales as $O(\varepsilon^{-1})$ and is independent of the code distance. To alleviate (ii) and enable efficient  high-dimensional optimization, we harness the sparse structure of the surrogate objective. 
We rely on the locality of detectors in the QEC circuit to narrow down the dependence of each term in $C$ to only a small subset of control parameters -- those related to the gates within the respective detecting region.
These relations can be conveniently represented using a factor graph, a bipartite graph in which the factor nodes correspond to detectors, and the variable nodes correspond to the system control parameters, see Fig.~\ref{fig2}(d) and Supp.~Mat.~II for more information. 

\subsection*{Learning algorithm}\label{methods:rl_algo}

Directly optimizing the surrogate objective $C$ is intractable for standard off-the-shelf optimizers. Although some algorithms \cite{scipy2020} are compatible with large-scale nonlinear optimization problems represented as sparse factor graphs, in practice additional considerations play an important role. Specifically, our objective function $C$ is highly stochastic, as it is obtained by averaging the binary detector values from a finite (and preferably small) number of QEC cycles. Furthermore, $C$ is non-stationary due to inherent system drift. Additionally, evaluating $C$ is expensive, since it requires updating the classical controller and acquiring data from the QEC detectors. Finally, in the future, $C$ must be optimized in real time during the quantum computation, favoring smooth and gradual improvements over disruptive large parameter steps that might lead to catastrophic consequences for the logical algorithm.

Given these properties, we adopted a multi-objective
policy-gradient reinforcement learning approach, in which signals from QEC detectors comprising $C$ serve as multiple objectives optimized simultaneously. 
Our learning algorithm is built on parameter-exploring policy gradients \cite{sehnke2010parameter}, an approach where an entire control policy is sampled in one piece, which provides a convenient interface with the classical controller. For robustness in the stochastic and non-stationary setting, we integrate several key techniques: proximal policy optimization \cite{schulman2017proximal} for stability, entropy regularization \cite{haarnoja2018soft} to maintain exploration, and a replay buffer \cite{mnih2015human} for improved sample efficiency. Crucially, the algorithm harnesses the sparse structure of the factor graph representation of $C$ for efficient variance reduction of the Monte Carlo gradient estimator \cite{mohamed2020monte} via gradient masking. The latter technique is adopted from Appendix~G of Ref.~\cite{sivak2024optimization}, where it was derived from the point of view of multi-agent learning. For completeness, a detailed exposition of our RL algorithm is provided in Supp. Mat. VIII.

\newpage
\onecolumngrid

\vspace{1em}
\begin{flushleft}
{\small Google Quantum AI and Google DeepMind}

\bigskip
{\small
\renewcommand{\author}[2]{#1$^\textrm{\scriptsize #2}$}
\renewcommand{\affiliation}[2]{$^\textrm{\scriptsize #1}$ #2 \\}

\newcommand{\xGoogle}{\affiliation{1}{Google Quantum AI}}

\newcommand{\xDeepMind}{\affiliation{2}{Google DeepMind}}

\newcommand{\xUMass}{\affiliation{3}{Department of Electrical and Computer Engineering, University of Massachusetts, Amherst, MA}}

\newcommand{\xUCSB}{\affiliation{4}{Department of Physics, University of California, Santa Barbara, CA}}

\newcommand{\Google}{1}
\newcommand{\DeepMind}{2}
\newcommand{\UMass}{3}
\newcommand{\UCSB}{4}

\author{Volodymyr Sivak}{\Google},
\author{Alexis Morvan}{\Google},
\author{Michael Broughton}{\Google},
\author{Rodrigo G.~Cortiñas}{\Google},
\author{Johannes Bausch}{\DeepMind},
\author{Andrew W. Senior}{\DeepMind},
\author{Matthew Neeley}{\Google},
\author{Alec Eickbusch}{\Google},
\author{Noah Shutty}{\Google},
\author{Laleh Aghababaie~Beni}{\Google},
% ...

\bigskip

\xGoogle
\xDeepMind
\xUMass
\xUCSB

}
\end{flushleft}

\twocolumngrid
\break
\newpage
\bibliographystyle{apsrevlongbib}
\bibliography{bib.bib}
    
\end{document}


\title{Supplementary Information\\``Reinforcement learning control of quantum error correction''}

    \author{Google Quantum AI and Google DeepMind}
    
    \maketitle

    \newpage
    \onecolumngrid
    \renewcommand\thefigure{S\arabic{figure}} 

    \setcounter{figure}{0}  
    \setcounter{table}{0}
    \maketitle
\tableofcontents
    \newpage
\section{Experimental setup}
\label{sec:device_info}

\subsection{Device parameters}

All our experiments were conducted on two Willow superconducting quantum processors \cite{acharya2024quantum}. The first one was only used for the distance-7 surface code as described in Section~\ref{sec:distance7}, and the second one for all other experiments involving distance-5 surface code and color code. 

\begin{figure}[h]
    \centering
    \includegraphics{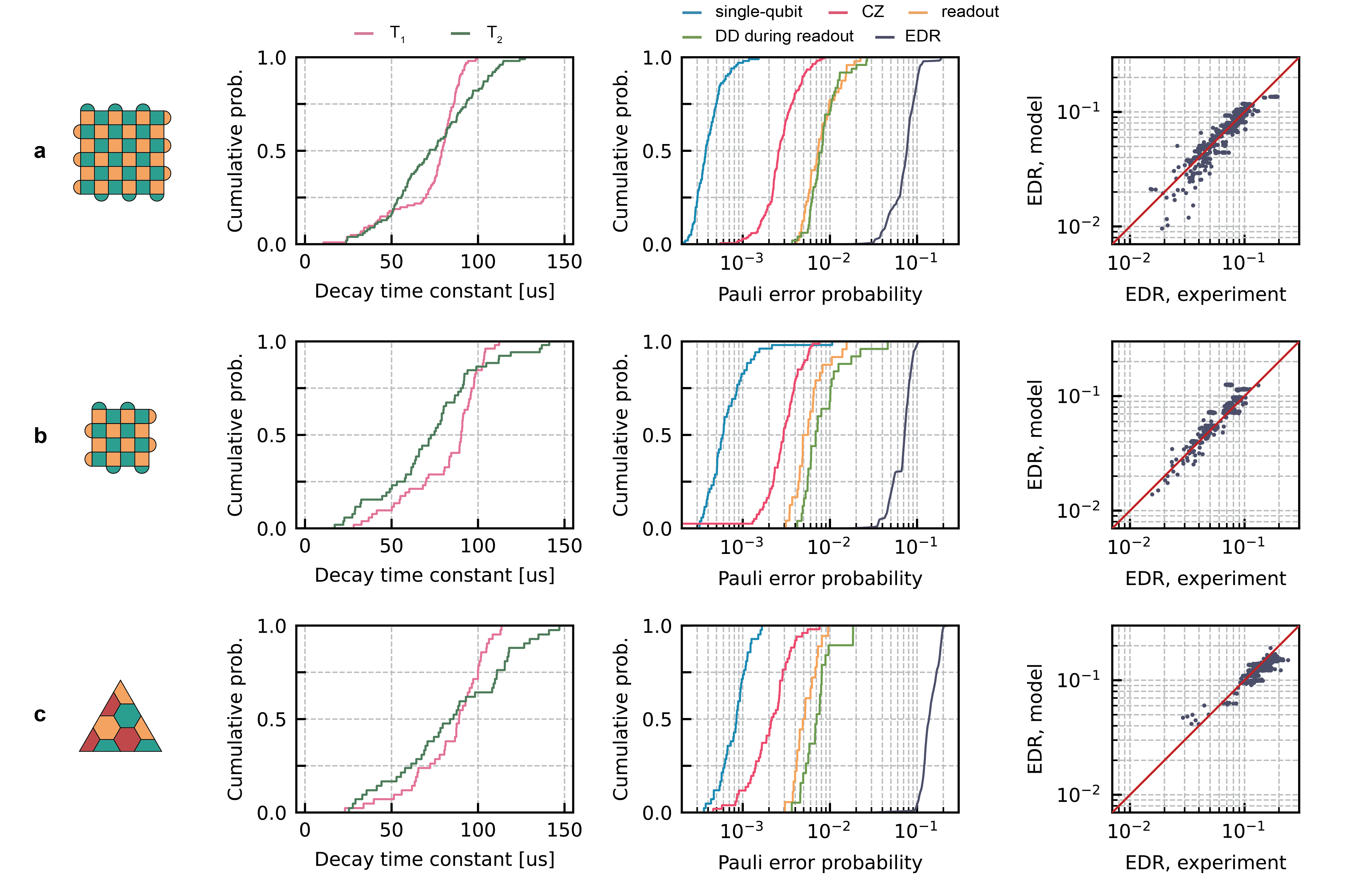}
    \caption{
    \label{fig_sm:benchmarks} 
    {\bf Device characterization.} 
    First column: cumulative distribution of $T_1$ and $T_2^{\rm CPMG}$ times across the qubits involved in the respective QEC circuits for {\bf (a)} distance-7 surface code, {\bf (b)} distance-5 surface code, and {\bf (c)} distance-5 color code.
    Second column: cumulative distribution of Pauli error rates for various gates, and of the error detection rates (EDR) for all detectors in the QEC circuit. 
    Third column: the EDR of all detectors, predicted from the Pauli simulation based on this characterization data, agrees well with the QEC experiment. Here the data is shown for experiments in both $X$ and $Z$ logical basis. 
    }
\end{figure}

Cumulative histograms of qubit relaxation times $T_1$, coherence times under dynamical decoupling $T_2^{\rm CPMG}$, as well as 
benchmarked Pauli error probabilities associated with different gates are shown in Fig.~\ref{fig_sm:benchmarks}. Note that these error probabilities drift over time, but the shown benchmarks are representative of typical performance. 
The Pauli model of the QEC process based on this characterization data, constructed with the methodology of Ref.~\cite{eickbusch2025demonstratingdynamicsurfacecodes}, agrees well with experiment for both surface and color code, see Fig.~\ref{fig_sm:benchmarks} last column.

\subsection{RL control settings \label{sm_sec:rl_settings}}

In a typical training run for the surface code, we sample $40$ policy candidates per epoch, which was empirically chosen as a compromise between wall clock runtime and learning efficiency, likely with room for improvement. To construct the reward vector for each candidate, we execute a $25$-cycle quantum memory experiment for $4 \times 10^3$ shots, resulting in $10^5$ effective QEC cycles per policy candidate. In a real-time steering scenario, this would correspond to approximately $0.1$ seconds of QEC per candidate (or $4$ seconds per epoch), given a typical $1\;\mu s$ cycle duration in superconducting circuits \cite{acharya2024quantum}. However, in our current setup, each training epoch lasts from $1$ to $10$ minutes.
These settings are not rigid; we sometimes adjust them to accommodate other considerations, such as compilation time. For example, in the color code experiments, we use $15$-cycle memory experiments instead of $25$-cycle.

After every five training epochs, we perform an evaluation epoch, during which we acquire $2 \times 10^5$ shots using both the learned policy $\mu(t)$ and the fixed initial policy $\mu(0)$. We re-evaluate the initial policy because its performance changes over time due to natural system drift. In contrast to the training data, this evaluation data is decoded to estimate the LER. The results in Fig.~3(a) show the relative improvement of LER measured during these evaluation epochs. The raw, unnormalized data from the same runs are presented and analyzed in Section~\ref{sec:multiple_runs}.

For the drift injection experiment shown in Fig.~4, we evaluate a third policy, $\mu'(0)$, which is identical to the initial policy $\mu(0)$ but does not have the artificial drift applied. The normalization in Fig.~4(a,c) then removes the effect of natural system drift by dividing the performance metrics (EDR or LER) of the $\mu(t)$ and $\mu(0)$ policies by the corresponding metric from the $\mu'(0)$ policy. The resulting trends reveal the learning behavior in response to the artificial drift alone.

    \section{Connecting control space to QEC detectors}
\label{sec:detectors} 

A central challenge in quantum control is navigating the vast, high-dimensional space of continuous parameters that define gate operations. Our approach leverages QEC detectors to guide this process, creating a direct link between the RL algorithm and the physical hardware. This section details this connection by first describing the tunable physical parameters that constitute the control space. We then describe how each detector is mapped to the subset of parameters influencing its outcome, a process that defines the sparse factor-graph structure of our optimization problem. Finally, we outline a rescaling procedure that normalizes this parameter space to ensure efficient learning. 

\subsection{Parametrization of control space}
\label{sec_sm:parametrization}
The complete set of tunable control parameters, which our reinforcement learning algorithm directly manipulates to optimize performance, is enumerated below:

\begin{itemize}
    \item \textbf{Single-qubit gates} are implemented via shaped microwave pulses using a raised cosine envelope with a Derivative Removal by Adiabatic Gate (DRAG) correction~\cite{Chow_2010, Chen_2016} to minimize leakage to higher energy states. The in-phase ($I$) and quadrature ($Q$) pulse components are defined by:
    \begin{align}
        \Omega_I(t) &= \frac{A}{2} \left[1 + \cos\left(\frac{2\pi t}{w}\right)\right],\quad -w/2 \leq t \leq w/2 \\
        \Omega_Q(t) &= -\frac{\alpha}{\Delta}\frac{d\Omega_I(t) }{dt},
    \end{align}
    where $w$ is the pulse width, $A$ is the pulse amplitude, $\Delta$ is the transmon anharmonicity, and $\alpha$ is the DRAG coefficient. The total complex pulse envelope, including a frequency detuning $\delta\omega = 2\pi f_d$, is given by 
    \begin{align}
        \Omega(t) &= (\Omega_I(t) + i\Omega_Q(t))\cdot e^{-i\delta\omega(t - w/2)}
    \end{align}
    The five control parameters for each single-qubit gate tuned by the RL agent include: $A$, $\alpha$, $f_d$, and additional virtual phase corrections before and after the pulse \cite{arute2019quantum}. 

    \item \textbf{CZ gates} are realized by varying the inter-qubit coupling strength $g(t)$ while simultaneously shifting the qubit frequencies via magnetic flux to bring the $|11\rangle$ and $|02\rangle$ states into resonance \cite{Foxen_2020}. We employ a smoothstep flattop pulse shape for the coupling trajectory, which consists of a smooth ramp-up, a plateau where the coupling is held constant at $g_{\text{max}}$, and a symmetric ramp-down. 
    The seven control parameters for each CZ gate tuned by the RL agent include: detuning of the coupler (controlling $g_{\text{max}}$), the detunings of both qubits during the gate, and additional virtual phase corrections on each qubit before and after the pulse \cite{arute2019quantum,Foxen_2020}.
        
     \item \textbf{Transfer functions} are used to apply pulse pre-distortions to correct for downstream spurious distortions in the control setup. We model the transfer function of each Z line to account for two primary effects ~\cite{hellings2025calibratingmagneticfluxcontrol}. First, the transient settling behavior is heuristically modeled with the transfer function ansatz from Ref.~\cite{jimmy_thesis}:
    \begin{equation}
        H_{\text{settling}}(\omega) = 1 + \sum_{k=1}^{N_{S}} \frac{a_{k} \cdot i\omega}{\omega_{S,k} + i\omega},
        \label{eq:settling_response}
    \end{equation}
    where for each of $N_S$ components, $a_k$ is the dimensionless amplitude (describing signal overshoot or undershoot) and $\omega_{S,k}$ is the exponential settling rate. Second, signal reflections from impedance mismatches are captured by the simple Fabry-Perot model \cite{pozar2021microwave}:
    \begin{equation}
        H_{\text{reflection}}(\omega) = \prod_{k=1}^{N_{R}} \frac{1-r_{k}}{1-r_{k}\exp(-i\,\omega / \omega_{R,k})},
        \label{eq:reflection_response}
    \end{equation}
    where for each of $N_R$ components, $r_k$ is the dimensionless amplitude and $\omega_{R,k}$ is the reflection rate. 
    
    We combine these two contributions, $H_{\text{settling}}(\omega)$ and $H_{\text{reflection}}(\omega)$, into the total transfer function and construct a pre-distortion filter $H_{\text{pre-distort}}(\omega)$ to be applied to the ideal control waveforms:
    \begin{equation}
        H_{\text{pre-distort}}(\omega) = H_{\text{settling}}^{-1}(\omega) \cdot H_{\text{reflection}}^{-1}(\omega)
        \label{eq:compensation_filter}
    \end{equation}
    All transfer function parameters $\{a_k, \omega_{S,k}\}_{k=1}^{N_S}$ and $\{r_k, \omega_{R,k}\}_{k=1}^{N_R}$ are included in the RL control space, and we choose $N_S=N_R=2$ for simplicity, which results in the total of eight control parameters of this kind per qubit.

    \item \textbf{Circuit phase corrections} allow to heuristically compensate for context-dependent coherent errors, such as crosstalk or residual detunings \cite{google2023suppressing,acharya2024quantum}. They are implemented by inserting virtual $Z^\beta$ rotations parametrized by the angle $\beta \in [-1, 1]$ into the circuit. These corrections are placed systematically before single-qubit gates and then simplified via commutation rules to remove redundancy. The standard calibration for these parameters is described in \cite{kelly2016scalable}. The rotation angles $\beta$ are included in our RL control space.

\end{itemize}

\begin{table}
    \centering
    \begin{tabular}{lc} 
        \toprule
        \textbf{Experiment} & \textbf{Policy size} \\
        \midrule
        $d=7$ surface code, fine-tuning experiment, Fig.~\ref{fig_sm:d7_training_run}  & 2657 \\
        $d=5$ surface code, fine-tuning experiment, Fig.~\ref{fig_sm:multiple_runs} &  1582 \\
        $d=5$ color code, fine-tuning experiment, Fig.~\ref{fig_sm:multiple_runs}  & 1489 \\
        $d=5$ surface code, randomized initial policy experiment, Fig.~\ref{fig_sm:randomized_initial_state} & 1582 \\
        $d=5$ surface code, drift injection experiment, Fig.~4  & 924 \\
        $d=3$ surface code, real-time steering simulation, Fig.~5(a)  & 41 \\        
        $d=15$ surface code, scaling simulation, Fig~5(b) & 38,670 \\
        \bottomrule
    \end{tabular}
    \caption{{\bf Number of learned control parameters in various experiments reported in this work.} \label{tab:policy_size}}
\end{table}

\subsection{Mapping detectors to control parameters}

\begin{figure*}[h]
    \centering
    \includegraphics{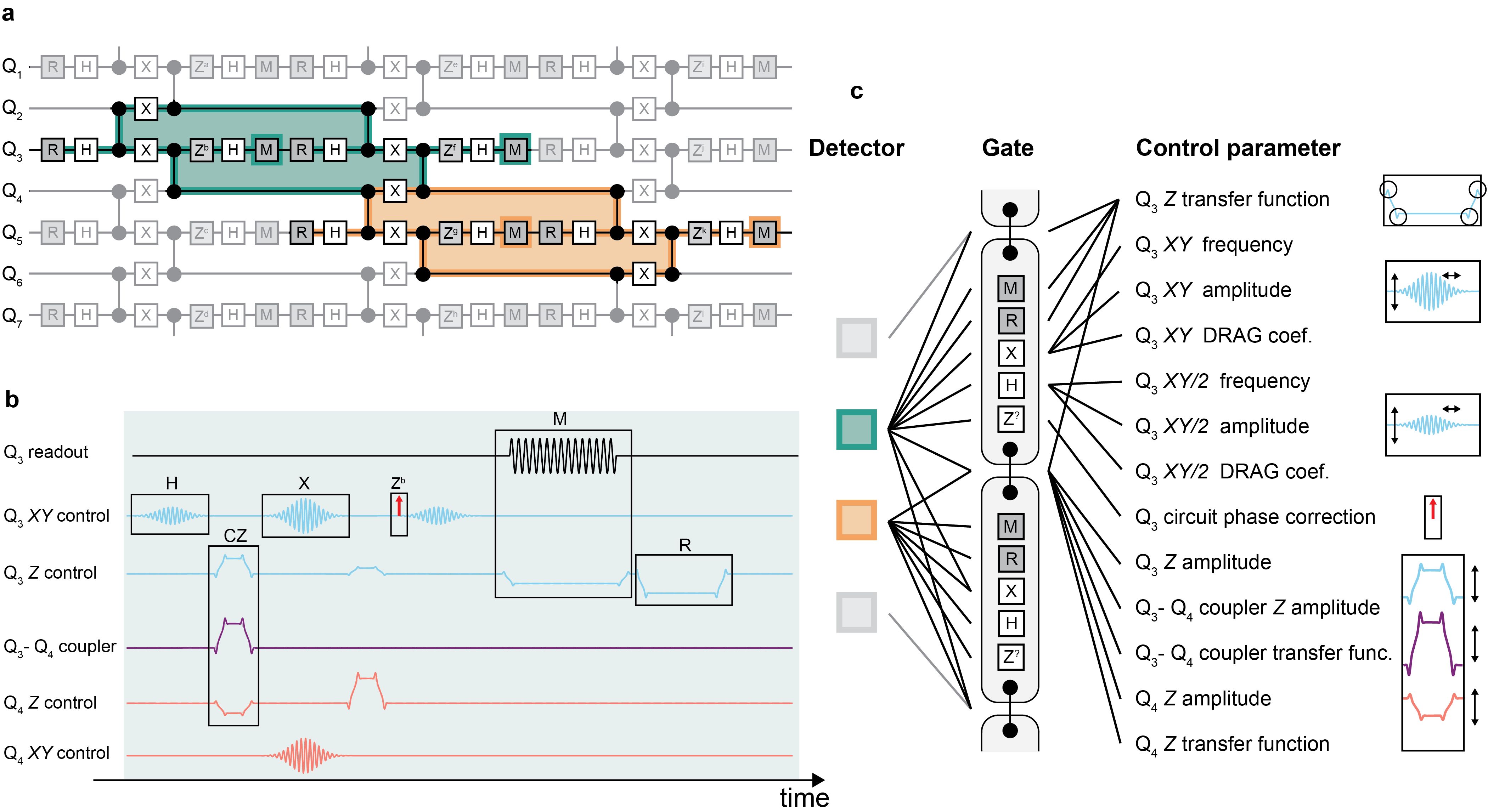}
    \caption{\label{fig_sm:rl_parameters} 
    \textbf{Mapping detectors to control parameters.}
    \textbf{(a)} Section of the repetition code circuit, similar to Fig.~2(a) but including additional virtual Z rotations. The shaded green and orange areas highlight the spacetime detecting regions for two distinct detectors.
    \textbf{(b)} Cartoon schematic of the analog control pulses that physically realize a portion of the gate sequence from (a). Abstract circuit operations, such as H, CZ, etc, are implemented with precisely shaped microwave and DC pulses.
    \textbf{(c)} For each  detector in (a), we identify the set of gates within its detecting region (center). The tunable control parameters for these gates (right), such as pulse amplitudes, frequencies, etc, collectively form the relevant parameter subspace for this detector.
    }
\end{figure*}

To improve the efficiency of our RL framework, we first reduce the dimensionality of the reward vector. We achieve this by combining detectors that are equivalent under time-translation into a single component of the reward vector \cite{sivak2024optimization}. This simplification reduces the size of the reward vector in a quantum memory experiment of duration $T$ from $O(d^2 T)$ to $O(d^2)$, where $d$ is the code distance. 

Next, we leverage the locality of detectors in a QEC circuit by utilizing gradient masking \cite{sivak2024optimization}, a technique that reduces the variance of the gradient estimator and ensures that the learning efficiency does not degrade with increasing system size. To implement this, we construct a mapping, illustrated in Fig.~\ref{fig_sm:rl_parameters}(c), from each detector to the gates within its detecting region, and subsequently from each gate to its constituent control parameters. This multi-step mapping is then simplified, resulting in a direct mapping from detectors to control parameters, which defines the sparse structure of our optimization problem, represented as a factor graph in Fig.~2(d).

The connectivity of the factor graph depends on a particular QEC algorithm, but in general it exhibits two characteristic features..
First, when several detectors share a common physical gate, the signal from all these detectors is used to learn the control parameters of that gate. Second, some low-level control parameters have a broad impact. For instance, the parameters of the transfer function of a Z line affect all operations driven by that line, including CZ gates, readout, and reset. Consequently, these parameters appear in the subspaces of multiple detectors. 

This structure results in a sparse but highly interconnected optimization problem. For example, in our distance-5 surface code experiment, the reward vector has $97$ components, the control space has $1582$ parameters, and on average each detector node is connected to $302$ parameter nodes, while each parameter node is connected to $18$ detector nodes.

\subsection{Calibration of detector sensitivities}

\begin{figure}[h]
    \centering
    \includegraphics{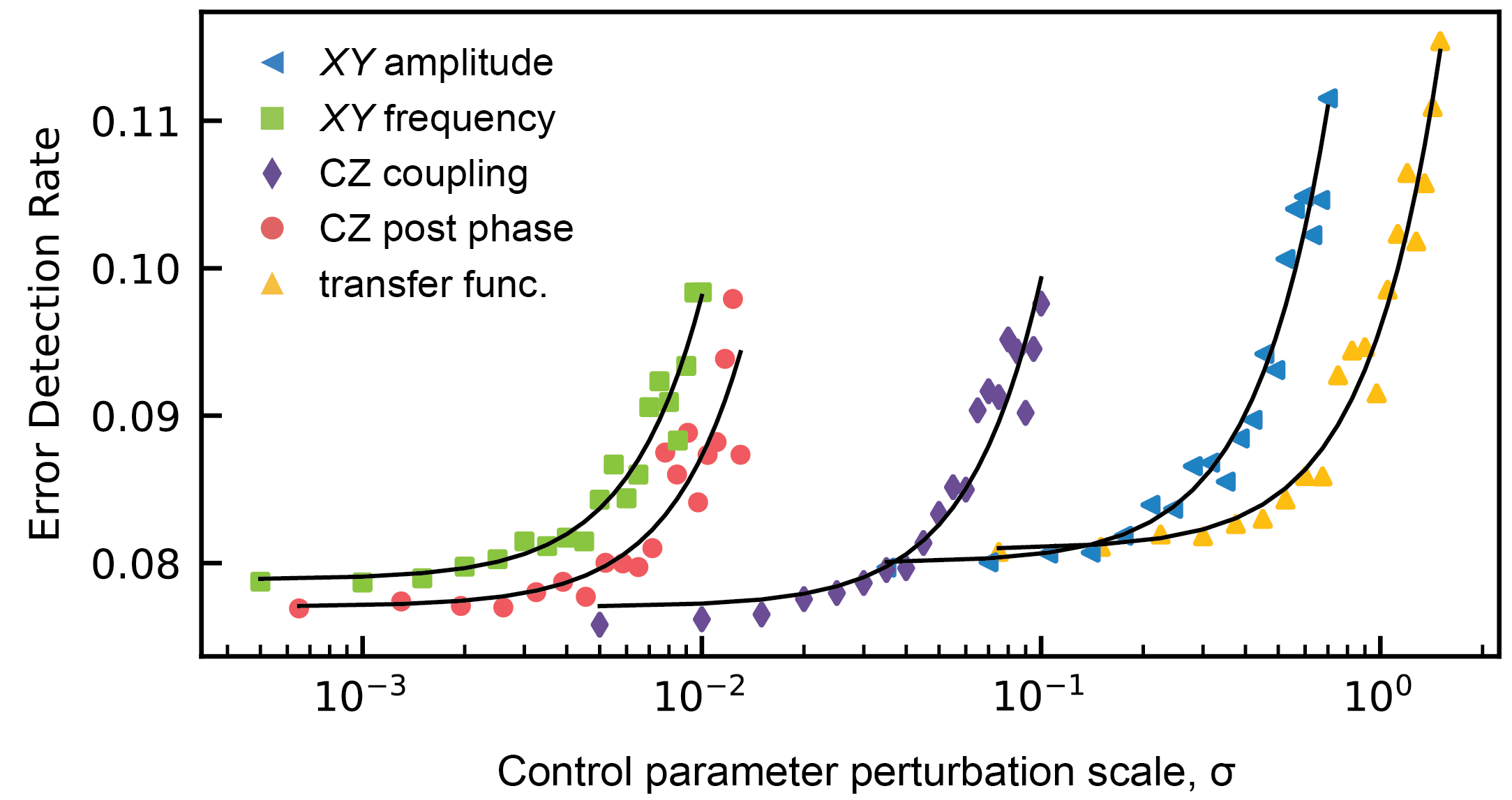}
    \caption{\label{fig_sm:detector_sensitivity} 
    {\bf Rescaling of control space.} Control parameters are typically defined over a wide numerical range, altogether spanning several orders of magnitude. To ensure that the Gaussian policy distribution is not strongly anisotropic, we rescale the control space in the following way. For each type of control parameter, e.g. XY amplitude, we sample perturbations from the distribution ${\cal N}(0, \sigma)$ and apply them to all gates that depend on this type of control parameter to imitate the exploration noise during the learning process. We sweep the variance $\sigma^2$ of the sampling distribution and measure the raise in error detection rate (EDR). By fitting this dependence to a quadratic model ${\rm EDR}={\rm EDR}_0 + (\sigma/\sigma_0)^2$ (black), we establish the coefficient $\sigma_0$, which quantifies the ``detector sensitivity'' to exploration noise in this type of control parameter. During the learning, we rescale all parameters by these sensitivity coefficients to ensure efficient behavior of the learning algorithm.
    }
\end{figure}

Our learning algorithm is most effective when the Gaussian policy distribution is not strongly anisotropic, partly due to architectural constraints such as using a single learning rate across all dimensions of the control space. In practice, however, the physical control parameters span several orders of magnitude, as seen in Fig.~\ref{fig_sm:detector_sensitivity}. A policy defined directly over this native parameter space would therefore be highly anisotropic, leading to an ill-conditioned optimization problem and hampering learning efficiency. 
To prevent this, we rescale all system control parameters in such a way that sampling the parameter perturbations from the canonical Gaussian $\delta p\sim {\cal N}(0,1)$ produces a comparable increase in the detection event rate regardless of the choice of axis $p$ in the control space.
This rescaling allows us to initialize the policy distribution in each RL run as spherically symmetric. In the later stages of the learning process the distribution inevitably deforms and becomes somewhat anisotropic as the algorithm self-regulates the amount of exploration and exploitation for each control parameter.

    \section{Analysis of multiple RL runs}
\label{sec:multiple_runs}

\begin{figure*}
    \centering
    \includegraphics{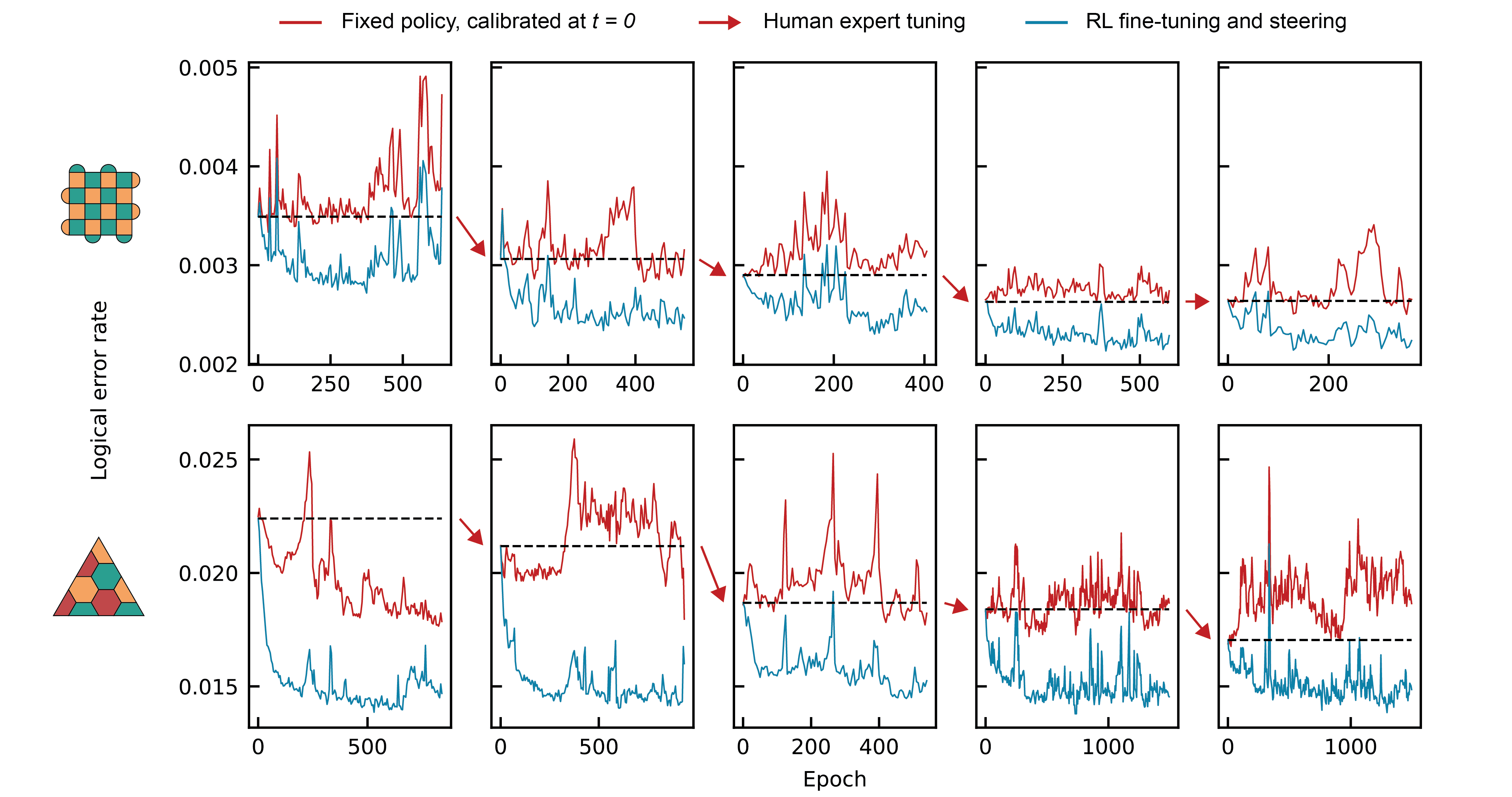}
    \caption{\label{fig_sm:multiple_runs}
    {\bf Multiple independent RL runs} for the distance-5 surface code (top) and color code (bottom). The LER values here are calculated using one-point estimate methodology, see Supp. Mat. Section VI in Ref. \cite{acharya2024quantum}. The LER improvement curves in Fig.~3(a) are obtained from these data.}
\end{figure*}

In Fig.~\ref{fig_sm:multiple_runs} we show multiple independent RL runs for the distance-5 surface and color codes. Each run starts after extensive calibration and human-expert tuning of the system. During these multi-day intervals (indicated by red arrows), human experts intervene to resolve edge-cases which are challenging to reliably automate.
Once this comprehensive tuning is finished, we launch the RL run and monitor its logical performance during the dedicated evaluation epochs (see Section~\ref{sm_sec:rl_settings} for additional details).
As explicitly shown in Fig.~3(a), RL systematically improves the logical performance by about $20\%$. It also significantly improves stability against natural drift, an effect we quantify next.

\begin{figure*}
    \centering
    \includegraphics{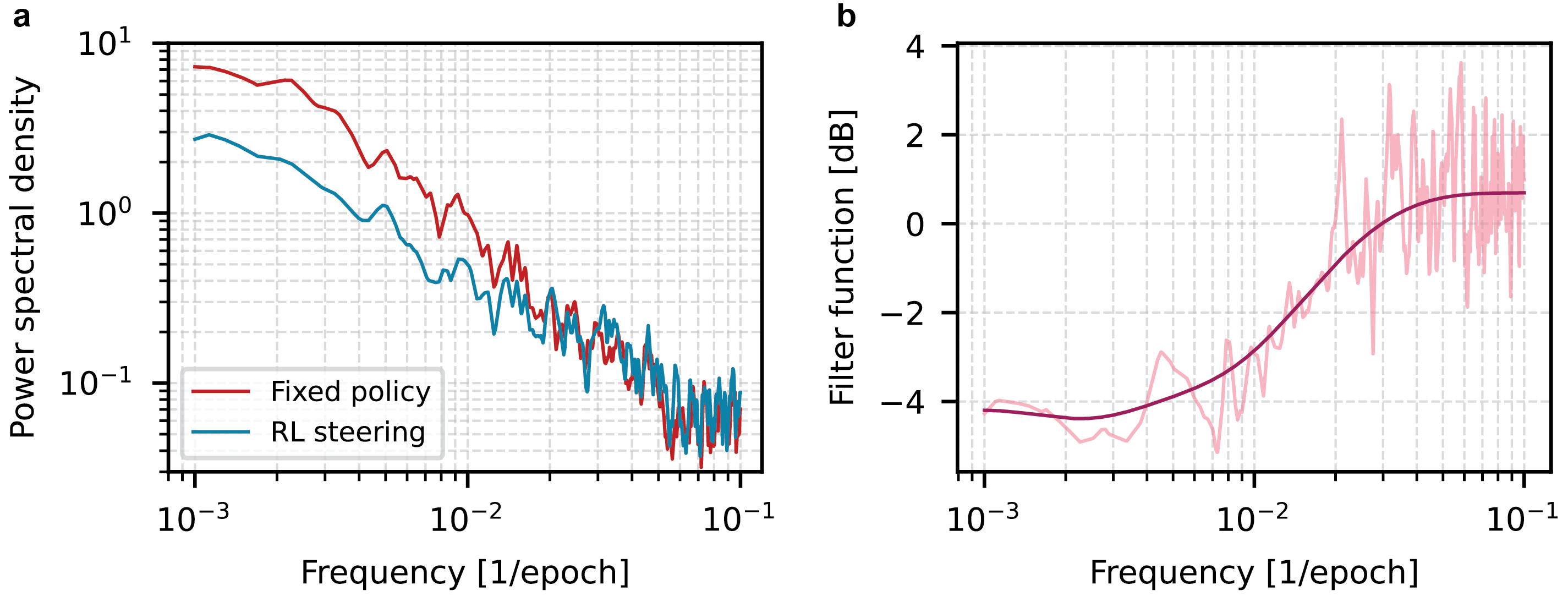}
    \caption{\label{fig_sm:power_spectrum}
    {\bf Analysis of LER fluctuations.} {\bf (a)} Average power spectral density of LER fluctuations from Fig.~\ref{fig_sm:multiple_runs}, see Section~\ref{sec:multiple_runs} for details. {\bf (b)} Effective filter function for LER fluctuations due to RL steering (pink) and guide to the eye (maroon) obtained with a smoothing Gaussian filter.
    }
\end{figure*}

We show the power spectral density (PSD) of LER fluctuations in Fig.~\ref{fig_sm:power_spectrum}(a) for both the fixed policy and RL steering. Note that the LER evaluations are done at regular 5-epoch intervals which are non-uniformly spaces in physical time (see discussion in Section~\ref{sm_sec:rl_settings}). Therefore, we perform the PSD analysis in the epoch domain to characterize fluctuations across the sequence of steering steps.  
To compute the average PSD, we first normalize the LER traces from independent runs in Fig.~\ref{fig_sm:multiple_runs}. We exclude initial 150 ``warmup'' epochs from the analysis, and the normalization ensures that all time traces start from value 1 at epoch 150. 
Then, we compute the power spectrum for each run using  discrete Fourier transform. Since all our RL runs have different lengths, we interpolate the individual spectra onto a shared frequency grid before computing their geometric average, excluding the zero-frequency component. This averaged data indicates that RL steering systematically reduces the amount of power at low frequencies by suppressing slow drift. The suppression is characterized using a filter function, shown in Fig.~\ref{fig_sm:power_spectrum}(b), which is empirically obtained by dividing the two spectra in Fig.~\ref{fig_sm:power_spectrum}(a). This filter function shows that low-frequency LER fluctuations are suppressed by approximately $4$ dB. Finally, note that treating RL steering using linear filter theory is not fully justified, but it provides useful physical intuition.

    \section{Performance with randomized initial policy}

\begin{figure*}
    \centering
    \includegraphics{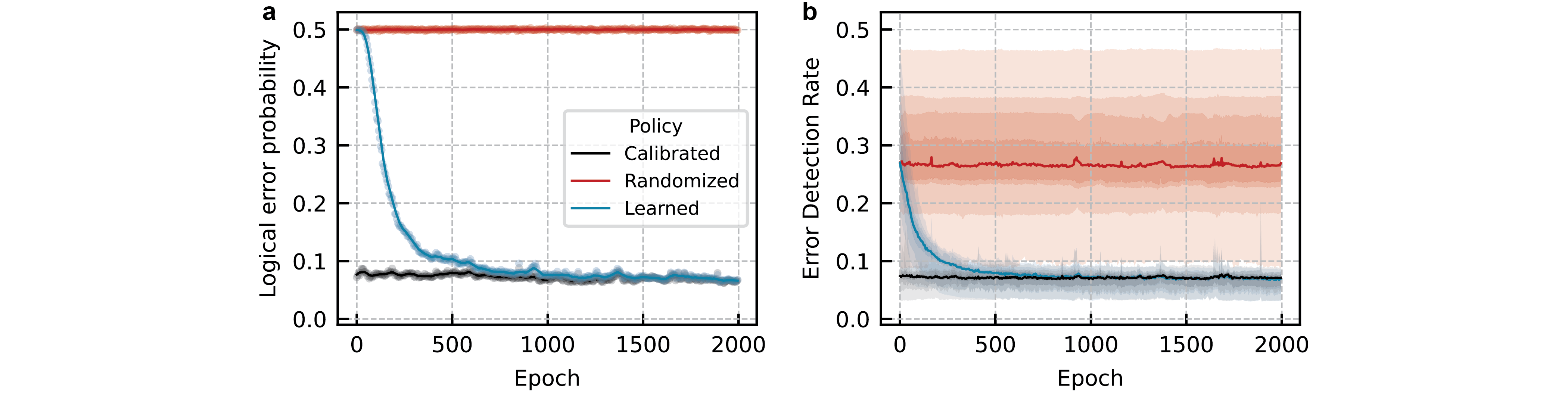}
    \caption{\label{fig_sm:randomized_initial_state}
    \textbf{RL training progress starting from a randomized initial policy.}
    \textbf{(a)} Evolution of the logical error probability of a 25-cycle surface code QEC circuit as a function of the training time. Initially, the ``randomized policy'' (red) has the logical error probability of $0.5$, which essentially corresponds to random guessing of the logical outcome. The ``learned policy'' (blue) starts from the randomized state and rapidly improves to match the performance of the expert-tuned ``calibrated policy'' (black). Circles show experimental data, and lines are guides to the eye obtained with a smoothing Gaussian filter.
    \textbf{(b)} Evolution of the error detection rate distribution across all detectors as a function of training time. Shaded regions represent quartiles of the data. 
    }
\end{figure*}

It is conceivable that in the future, with sufficient enhancement of our learning framework, a quantum processor could be calibrated for QEC {\it ab initio} fully by RL, with no reliance on the traditional calibration paradigm or human experts. 
While this requires significant enhancement of our RL framework, here we emulate a simpler but related scenario of a coarsely calibrated device. 
We start the RL training from a randomized initial policy. The randomization is achieved by first calibrating the device for the distance-5 surface code, and then deliberately spoiling all control parameters described in Section~\ref{sec_sm:parametrization} by a sufficient amount to bring the logical error probability of the full 25-cycle memory circuit to the $50\%$ level (fully random logical outcome). 
In contrast to the training results in Section~\ref{sec:multiple_runs}, where RL starts from a well-calibrated state and is able to make significant progress in just about 200 epochs, the current scenario puts the RL agent into more challenging conditions. As a result, it takes about $1000$ epochs to recover the performance to the level of the calibrated policy, see Fig.~\ref{fig_sm:randomized_initial_state}. Nevertheless, this experiment demonstrates that RL agent is able to bridge large gaps in logical performance. 

    \section{Distance-7 surface code experiment}
\label{sec:distance7}

\begin{figure*}
    \centering
    \includegraphics{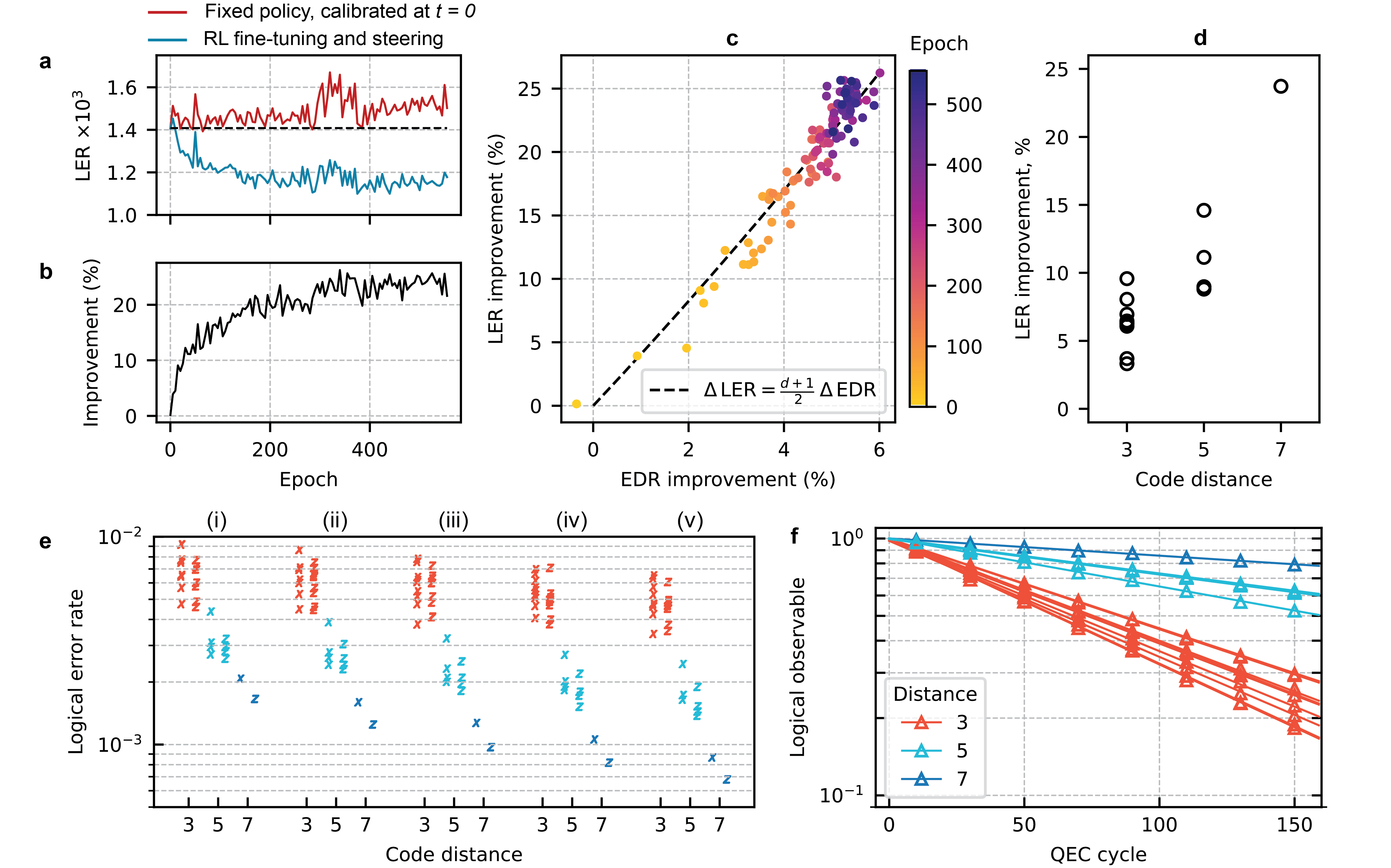}
    \caption{\label{fig_sm:d7_training_run}
    {\bf RL fine-tuning run for the distance-7 surface code.} 
    {\bf (a)} Evolution of the logical error rate during the RL run. The LER is calculated using one-point estimate from 25-cycle QEC memory circuit, see Supp. Mat. Section VI in Ref. \cite{acharya2024quantum}.
    {\bf (b)} Relative improvement of LER over time, extracted from (a).
    {\bf (c)} Correlation between the relative improvements in LER and EDR follows the expected trend based on the simple surface code scaling model (main text).
    {\bf (d)} Applying the learned control policy to smaller-distance codes located within the original distance-7 grid improves their LERs, demonstrating successful generalization.
    {\bf (e)} Successive last-mile performance improvements segregated by code distance and logical basis (marker shape). (i) Traditional calibration of the system and human-expert tuning; the decoding is done with Sparse Blossom decoder and SI1000 prior; (ii) Improved system calibration by adding RL fine-tuning; (iii) Switching to Tesseract decoder with SI1000 prior; (iv) Tesseract decoder with frequency-calibrated prior, see Section~\ref{sec:tesseract}; (v) Switching to AlphaQubit2 neural network decoder, fine tuned on an independent dataset with similar characteristics, see Section~\ref{sec:aq2}.
    {\bf (f)} Logical performance evaluation with AlphaQubit2 decoder, averaged over $X$ and $Z$ logical bases for each subgrid.    
    \label{figSM:d7_training}
    }
\end{figure*}

Our distance-7 surface code experiment was done on a different Willow processor from the rest of the experiments reported here; its performance is summarized in Fig.~\ref{fig_sm:benchmarks}. Aside from a different processor, we followed the same methodology as in the rest of this work, with results of the RL fine-tuning run summarized in Fig.~\ref{fig_sm:d7_training_run}(a-b). We find that the improvement of the LER is correlated with the improvement of the EDR via a trend expected from the simple scaling model (see main text): $\Delta {\rm LER}=\frac{d+1}{2}\Delta {\rm EDR}$, where $\Delta$ indicates the relative change, see Fig.~\ref{fig_sm:d7_training_run}(c).

While the training used a fixed distance-7, $Z$-basis, 25-cycle memory circuit, we evaluated the logical performance at other memory durations (sweeping from 10 to 150 cycles), logical bases ($X$ and $Z$), code distances (3, 5, 7), and subgrids within a larger distance-7 grid. For this evaluation, we used the control policy from epoch 0 (initial policy, calibrated via the traditional approach) and epoch 220 (learned policy). We find that the logical error probability of all evaluated circuits is reduced when using the learned policy, demonstrating successful generalization, see Fig.~\ref{fig_sm:d7_training_run}(d). However, we do not report the error suppression factor $\Lambda$ as in Ref.~\cite{acharya2024quantum}, because RL fine-tuning was done only on the distance-7 code, which might artificially bias $\Lambda$ towards a higher value. 

The logical error rates of all subcodes, obtained with different decoders, are shown in Fig.~\ref{figSM:d7_training}(e) and summarized in Table~\ref{tab:decoder_error_rates}. The logical decay curves with our most accurate decoder AlphaQubit2 are presented in Fig.~\ref{figSM:d7_training}(f).

\begin{table*}[htbp]
\centering
\renewcommand{\arraystretch}{1.3} 
\setlength{\tabcolsep}{8pt} 
\begin{tabular}{|c|l|cc|cc|cc|}
\hline
\multirow{2}{*}{Distance} & \multicolumn{1}{c|}{\multirow{2}{*}{Grid}} & \multicolumn{2}{c|}{Sparse Blossom, SI1000} & \multicolumn{2}{c|}{Tesseract} & \multicolumn{2}{c|}{AlphaQubit2} \\ \cline{3-8} 
 & & $X$ basis & $Z$ basis & $X$ basis & $Z$ basis & $X$ basis & $Z$ basis \\ \hline
\multirow{9}{*}{3} & (0, 0) & 0.599(6)\% & 0.537(2)\% & 0.525(3)\% & 0.469(4)\% & 0.477(4)\% & 0.451(4)\% \\
 & (0, 4) & 0.709(12)\% & 0.708(17)\% & 0.613(8)\% & 0.499(16)\% & 0.572(7)\% & 0.467(15)\% \\
 & (0, 8) & 0.711(10)\% & 0.667(8)\% & 0.553(10)\% & 0.486(5)\% & 0.526(9)\% & 0.465(5)\% \\
 & (4, 0) & 0.693(18)\% & 0.567(5)\% & 0.573(11)\% & 0.487(3)\% & 0.652(12)\% & 0.483(4)\% \\
 & (4, 4) & 0.448(3)\% & 0.751(4)\% & 0.406(3)\% & 0.708(3)\% & 0.340(18)\% & 0.606(3)\% \\
 & (4, 8) & 0.621(3)\% & 0.445(3)\% & 0.507(3)\% & 0.3795(16)\% & 0.462(2)\% & 0.352(2)\% \\
 & (8, 0) & 0.865(13)\% & 0.641(4)\% & 0.698(10)\% & 0.511(3)\% & 0.652(12)\% & 0.483(4)\% \\
 & (8, 4) & 0.864(5)\% & 0.650(3)\% & 0.664(4)\% & 0.533(3)\% & 0.630(5)\% & 0.495(3)\% \\
 & (8, 8) & 0.530(4)\% & 0.460(2)\% & 0.464(3)\% & 0.3992(18)\% & 0.424(3)\% & 0.377(2)\% \\ \hline
\multirow{4}{*}{5} & (0, 0) & 0.242(4)\% & 0.242(4)\% & 0.183(3)\% & 0.172(3)\% & 0.164(2)\% & 0.153(3)\% \\
 & (0, 4) & 0.2615(15)\% & 0.2306(11)\% & 0.1901(17)\% & 0.1520(9)\% & 0.173(2)\% & 0.144(11)\% \\
 & (4, 0) & 0.388(5)\% & 0.304(2)\% & 0.270(4)\% & 0.2192(13)\% & 0.173(2)\% & 0.138(8)\% \\
 & (4, 4) & 0.2777(12)\% & 0.2599(16)\% & 0.2015(14)\% & 0.1795(12)\% & 0.244(3)\% & 0.190(15)\% \\ \hline
7 & (0, 0) & 0.1599(14)\% & 0.1249(8)\% & 0.1059(13)\% & 0.0818(8)\% & 0.0866(11)\% & 0.0679(10)\% \\ \hline
\end{tabular}
\caption{\textbf{Logical error rates} for the QEC dataset obtained with RL fine-tuned control policy, using different decoders corresponding to columns (ii), (iv) and (v) in Fig.~\ref{figSM:d7_training}(e).}
\label{tab:decoder_error_rates}
\end{table*}
    \section{Simulation details}
\label{sec:sim}

\subsection{Real-time steering \label{sec_sm:steering}}

To capture the essential features of real-time steering, we designed a simple drift simulation of a distance-3 surface code in Stim \cite{gidney2021stim}. In this simulation, we include one learnable control parameter, $p_i$, for every single-qubit and two-qubit gate, where $i$ enumerates the gates. All parameters are subject to sinusoidal drift at a shared frequency $f$, such that the optimal parameter value at time $t$ evolves as $p_i^{\rm opt}(t) = \sin(2\pi ft)$. We model the depolarization rate for each gate as $\varepsilon_i(t, p_i)=\tilde{\varepsilon}_i+\Omega_i(p_i-p_i^{\rm opt}(t))^2$, where $\tilde{\varepsilon}_i$ is an irreducible error rate and the ``sensitivity'' $\Omega_i$ is chosen randomly for each gate. The optimal control policy is therefore to track the drift, $p_i(t)=p_i^{\rm opt}(t)$, while a fixed policy, $p_i(t)\equiv0$, is optimal only at $t=0$.

The simulation, featured in Fig.~5(a), executes a total of $1.8\times10^9$ QEC cycles, divided into $10^3$ training epochs. Following the procedure in Fig.~2(b), every epoch consists of a batch of $50$ policy candidates, with each candidate taking $3.6\times10^4$ QEC cycles. We then evaluate the performance by counting the total number of error detection events, $N$, generated under four distinct scenarios:
\begin{enumerate}[label=(\roman*)]
    \item a fixed policy $\mu(0)$, emulating a system calibrated at $t=0$; 
    \item the known optimal policy at all times;
    \item the ``stochastic policy'', which represents true real-time steering where QEC cycles are executed using the sampled policy candidates; 
    \item the learned policy $\mu(t)$, using the mean of the policy distribution (instead of sampling) at each epoch $t$.
\end{enumerate}

\begin{figure*}
    \centering
    \includegraphics{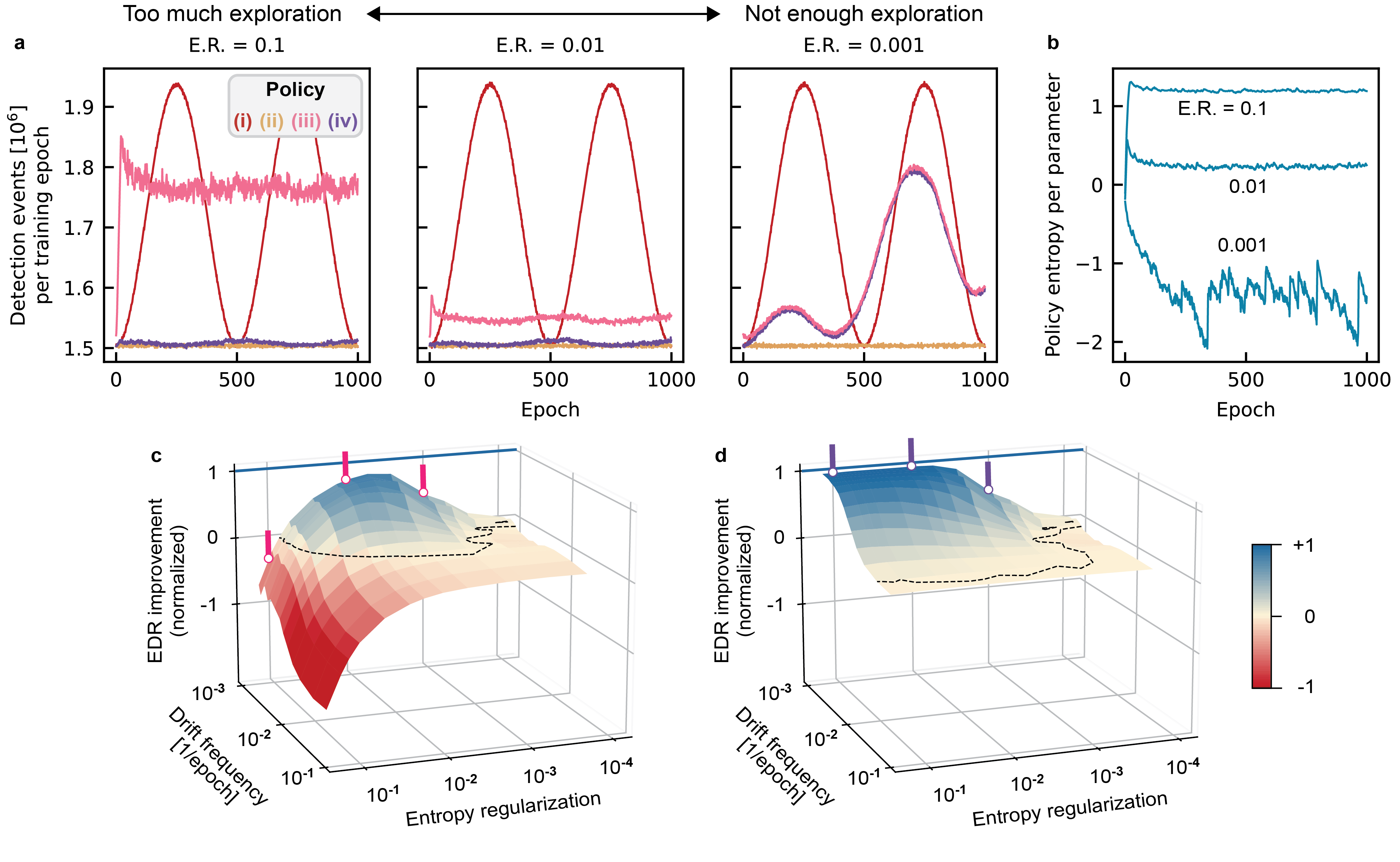}
    \caption{\label{fig_sm:steering_sim}
    {\bf Real-time steering simulation.} 
    {\bf (a)} Evolution of detection event rate over time in a drift simulation with drift frequency $f=1/(1000\; \rm epochs)$, using miscellaneous policies: (i) fixed, (ii) optimal, (iii) stochastic, (iv) learned, and different amounts of entropy regularization (E.R.). When exploration and exploitation are well-balanced (middle panel), the performance of real-time steering (i.e. stochastic policy) approaches close to optimal. 
    {\bf (b)} Evolution of the average policy entropy over time; higher entropy corresponds to more exploration.
    {\bf (c)} Normalized improvement of the cumulative error detection rate (EDR) of the stochastic policy over the fixed policy for different simulation parameters. Values greater than 0, separated by the dashed borderline, correspond to performance better than the fixed policy. Pink notches indicate simulation parameters chosen for (a). 
    {\bf (d)} Same as (c), but for the learned policy instead of the stochastic policy. See Section~\ref{sec_sm:steering} for additional details.
    }
\end{figure*}

The evolution of the detection rates for these four scenarios, shown in Fig.~\ref{fig_sm:steering_sim}(a), depends strongly on the exploration-exploitation balance, which we control via an entropy regularization (E.R.) hyperparameter. For ${\rm E.R.}=0.1$, excessive exploration noise degrades the performance of the stochastic policy, even though it enables the learned policy $\mu(t)$ to efficiently track the optimum. For ${\rm E.R.}=0.01$, the tradeoff is well-balanced, and the performance of the stochastic policy approaches the optimum. Finally, for ${\rm E.R.}=0.001$, insufficient exploration prevents the learning agent from keeping pace with the drift. This behavior is consistent with the average policy entropy shown in Fig.~\ref{fig_sm:steering_sim}(b), which is highest for the exploratory regime and lowest for the exploitative one.

To quantify this tradeoff more systematically, we scan the drift frequency and entropy regularization, as shown in Fig.~\ref{fig_sm:steering_sim}(c) and Fig.~5(a). We compute the ratio $r=(N_{(iii)} - N_{(i)})/(N_{(ii)}-N_{(i)})$ to measure the real-time steering advantage. A value of $r=1$ indicates maximally efficient steering that matches the optimal policy, while $r \le 0$ indicates that the steering performs worse than the fixed policy due to excessive exploration noise. As explained in the main text, this analysis reveals a ``steerability threshold'' -- a maximum drift frequency of about $1/(150\; \rm epochs)$ above which real-time steering cannot keep up. Below this frequency, exploration and exploitation can be balanced to realize a significant performance advantage from RL steering, indicated by the $r>0$ region in the phase diagram.

Finally, it is crucial to distinguish between the performance of the stochastic policy and the learned policy $\mu(t)$. As shown in Fig.~\ref{fig_sm:steering_sim}(d), the learned policy outperforms the fixed policy over a much larger region of the phase diagram. This is because the performance of $\mu(t)$ is not penalized by the exploration noise inherent to the stochastic policy. All experimental results in our work use the learned policy $\mu(t)$ to quantify the advantage of RL control. However, for future single-shot logical algorithms, the performance of the stochastic policy will become a more relevant metric, as exploration noise will directly impact the computation under real-time steering.

\subsection{Scaling to large code distance}
\label{sec:scaling_supp}

In the scaling study featured in Fig.~5(b), we simulate a $10$-cycle QEC memory experiment in Stim for surface codes of odd distances $d=3-15$. 
We model the depolarization rate for each single-qubit and two-qubit gate as $\varepsilon_i=\tilde{\varepsilon}_i+\sum_j\Omega_{i,j}\,p^2_{i,j}$, where $j=1,...,P$ enumerates the $P$ control parameters per gate, $\Omega_{i,j}$ is the (randomly chosen) sensitivity of the error rate of gate $i$ to miscalibration of parameter $j$, and $\tilde{\varepsilon}_i$ is the irreducible error rate.
The total number of learned parameters is given by
\begin{align}
    P_{\rm tot}=\underset{\rm num.\; 1q\; gates}{(2d^2-1)}\cdot P + \underset{\rm num.\; 2q\; gates}{(4d^2-4d)}\cdot P,
\end{align}
which for our largest simulation with $d=15$ and $P=30$ corresponds to $P_{\rm tot}=38,670$.

Each RL run is initialized with control parameters $p_{i,j}$ chosen uniformly at random from an interval sized to ensure that initial QEC performance is {\it above} the surface code threshold. The RL agent then progressively drives the parameters towards their optimal values at $p_{i,j}^{\rm opt}=0$, pushing the error rates $\varepsilon_i$ towards their irreducible values $\tilde{\varepsilon}_i$. Throughout this process, we compute the average physical error rate for any given policy by constructing its detector error model in Stim and averaging the error probabilities within it. Using this method, we located the threshold in Fig.~5(b) at a physical error rate of $1.79\cdot10^{-3}$.

To analyze the convergence dynamics, we construct point-estimates of the error suppression factor $\Lambda$ using a simple scaling model for the logical error rate $\varepsilon_{\rm L} = C\cdot\Lambda^{-(d+1)/2}$, which yields the relation $\Lambda = \Lambda^*\cdot(\varepsilon_{\rm L}^*/\varepsilon_{\rm L})^{2/(d+1)}$. Applying this to the data from Fig.~5(b) reveals that the learning speed near the optimum satisfies the empirical relation:
\begin{align}
    \partial_t\Lambda/\Lambda^* = \gamma\cdot (1-\Lambda/\Lambda^*)
    \label{eq:diff_eq_lambda}
\end{align}
Crucially, the convergence rate $\gamma$ in this relation is independent of the system size but depends on local properties, such as the number of gates per detecting region and the number of learnable parameters per gate, see Fig.~5(c). Solving this differential equation yields the exponential convergence law for $\Lambda$ near the local optimum:
\begin{align}
    (\Lambda^*-\Lambda)/\Lambda^*\propto e^{-\gamma\, t}
\end{align}
This favorable convergence is a direct result of our learning algorithm's ability to effectively utilize the sparsity of the optimization factor graph.
    \section{Decoding of QEC data}
\label{sec:decoding}

In this section, we provide information about QEC decoding methods. Comparison of our QEC results to previously published results is provided in Tables~\ref{tab:published_results}, \ref{tab:best_qec_results}, \ref{tab:results_with_equiv_decoding} and Fig.~\ref{fig_sm:google_history}.

The choice of decoder creates an additional source of variance when comparing data from different experiments. To eliminate this variance, in Table~\ref{tab:results_with_equiv_decoding} we summarize the results obtained with the same decoding method across different publicly released datasets. In this comparison, surface code is decoded with Sparse Blossom \cite{higgott2023sparse} and color code with Tesseract \cite{beni2025tesseract}, both using SI1000 decoder prior \cite{gidney2021fault}. The hyperparameters of Tesseract in this comparison were set to \code{no-revisit-dets=True}, \code{beam-climbing=True}, \code{pqlimit=1e5}, \code{det-penalty=0}, \code{beam=10}.

\subsection{Sparse Blossom}
\label{sec:sp_decoder}

Most of the QEC decoding in this work was done with a fast Sparse Blossom decoder  and a generic SI1000 decoder prior; this includes results in Figs.~2(c), 3(a), 4(c), 5(b), \ref{fig_sm:multiple_runs}, \ref{fig_sm:randomized_initial_state}, and \ref{figSM:d7_training}. 

Decoder steering in Fig.~4(c) used this decoder as well. We re-trained the parameters of the detector error model, which serves as a decoder prior, in between the evaluation epochs. This training was done with the RL method from Ref.~\cite{sivak2024optimization} and was completely decoupled from the main RL training of the system control parameters; it ran after the completion of the steering experiment using the saved QEC data. 

High-performance decoding of the main QEC datasets was done with Tesseract \cite{beni2025tesseract} and AlphaQubit2 \cite{senior2025scalable} decoders, described in the next sections.

\subsection{Tesseract}
\label{sec:tesseract}

The hyperparameters of Tesseract for decoding the surface code data (results summarized in Table~\ref{tab:decoder_error_rates}) were set to:
\\\code{no-revisit-dets=True}, \code{beam-climbing=True}, \code{det-order-index=True}, \code{det-penalty=0}, \code{beam=10}, \code{pqlimit=1e7}. 
For the color code, we used \code{pqlimit=1e5} instead. With these parameters, Tesseract’s runtime ranged from $45\;\rm \mu s$ to $27\;\rm ms$ per QEC cycle. 

We calibrated the detector error model (DEM) using Tesseract’s frequency calibration infrastructure \cite{tesseract_repo} which leverages the same technique as was used in Ref.~\cite{lacroix2025scaling} with an integer program decoder. This calibration is done separately for each code distance, logical basis ($X$ or $Z$), and qubit subgrid (involving a subset $S$ of all physical qubits). For each such subset $S$, there are multiple datasets corresponding to different numbers of QEC cycles. To avoid any potential overfitting, we ensure that the DEM used for the final decoding of one dataset is calibrated exclusively by decoding {\it other} datasets. To implement this exclusion, we partition the datasets into type A and type B cycles in an alternating sequence. We use Tesseract to decode all type A (B) data to a predicted set of most likely errors. We aggregate the total number of times each error mechanism was predicted across all shots within type A (B) cycles, and further aggregate errors corresponding to the same error mechanism time-translated over the QEC cycles in the DEM. We exclude shots with logical errors from the totals, then use the empirical rate of occurrence of each physical error mechanism to set its updated probability across all occurrences in the DEM.  We repeat this process for 4 iterations in the surface code and 5 iterations in the color code. Finally we generate a calibrated DEM for type B (A) cycles and decode once in an evaluation run on type B (A) data.

\begin{figure*}
    \centering
    \includegraphics{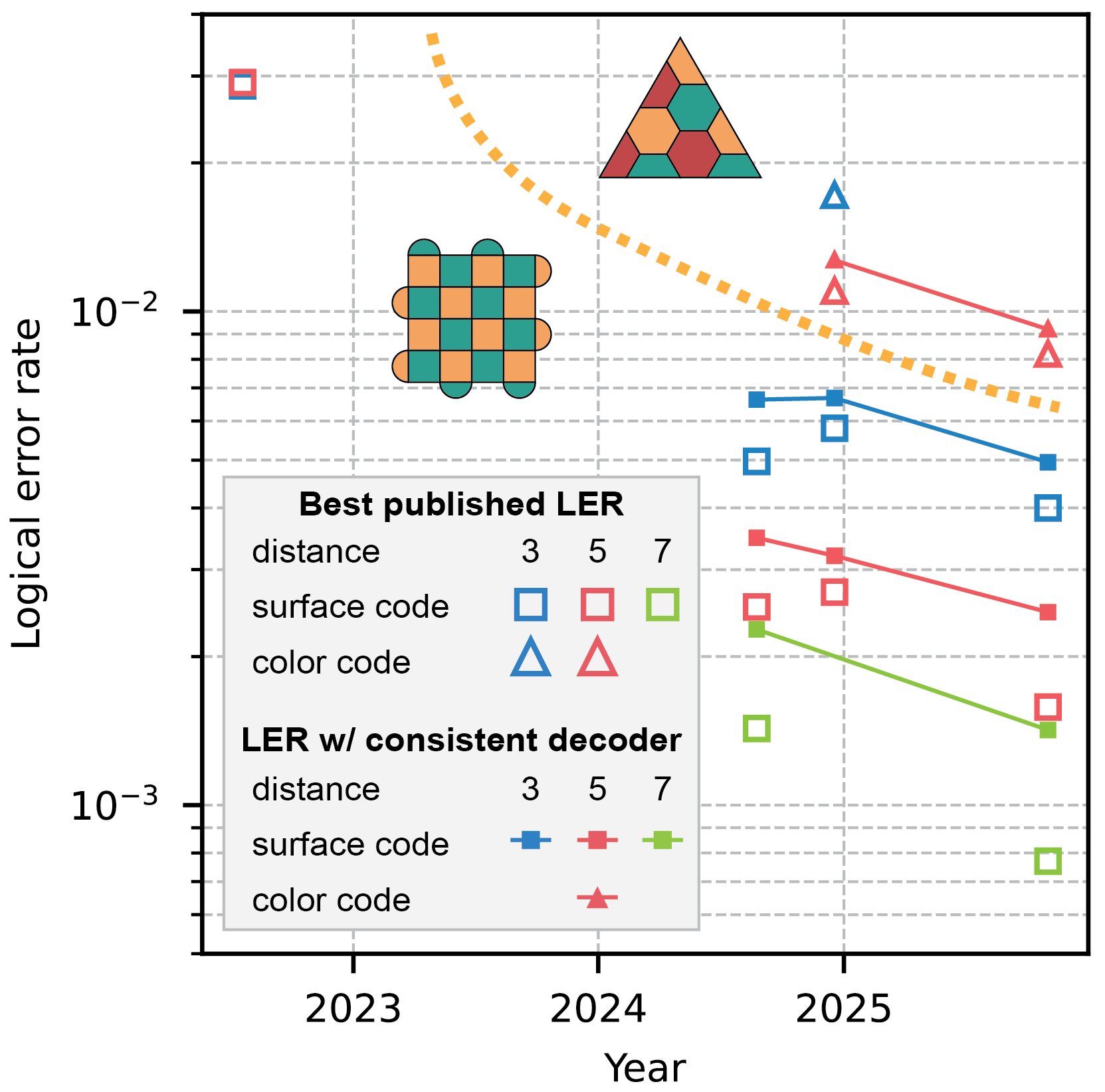}
    \caption{\label{fig_sm:google_history}
    {\bf Published QEC memory results from the Google Quantum AI team:} surface code (Refs. \cite{google2023suppressing}, \cite{acharya2024quantum}, \cite{eickbusch2025demonstratingdynamicsurfacecodes}, and this work) and color code (Ref. \cite{lacroix2025scaling} and this work). Empty symbols show the best published results, obtained with the most accurate decoders in the respective references, see Table~\ref{tab:best_qec_results}. Solid symbols connected with lines show results obtained for the same QEC datasets under consistent decoding conditions, see Table~\ref{tab:results_with_equiv_decoding}. The orange dashed curve separates color code and surface code experiments. The date for each experiment is chosen as the date of the first arXiv posting.
    }
\end{figure*}

\begin{table}
    \centering
    \begin{tabular}{ll} 
        \toprule
        \textbf{Publication} & \textbf{Link to data} \\
        \midrule
        Suppressing quantum errors by scaling a surface code logical qubit \cite{google2023suppressing} & \href{https://zenodo.org/records/6804040}{zenodo.6804040} \\
        Quantum error correction below the surface code threshold \cite{acharya2024quantum} & \href{https://zenodo.org/records/13273331}{zenodo.13273331} \\
        Demonstration of dynamic surface codes \cite{eickbusch2025demonstratingdynamicsurfacecodes}  & \href{https://zenodo.org/records/14238907}{zenodo.14238907} \\
        Scaling and logic in the color code on a superconducting quantum processor \cite{lacroix2025scaling}  & \href{https://zenodo.org/records/14238944}{zenodo.14238944} \\
        Reinforcement learning control of quantum error correction & \href{https://zenodo.org/records/17566522}{zenodo.17566522} \\
        \bottomrule
    \end{tabular}
    \caption{{\bf Published QEC memory datasets from the Google Quantum AI team.} \label{tab:published_results}}
\end{table}

\begin{table}
    \centering
    \renewcommand{\arraystretch}{1.3} 
    \setlength{\tabcolsep}{6pt}       
    \begin{tabular}{|c|c|c|c|c|c|}
        \hline
        \textbf{Code} & \textbf{Distances} & \textbf{LER} & \textbf{Decoder} & \textbf{Reference} & \textbf{Source} \\ 
        \hline \hline 
        \multirow{2}{*}{Surface code} & 3 & $2.87 \times 10^{-2}$ & \multirow{2}{*}{Tensor network} & \multirow{2}{*}{\cite{google2023suppressing}} & Fig. S32, top right subgrid \\
                                      & 5 & $2.91 \times 10^{-2}$ &                                 &                                               & Fig. S30  \\ \hline
        \multirow{3}{*}{Surface code} & 3 & $4.97 \times 10^{-3}$ & \multirow{3}{*}{Neural network} & \multirow{3}{*}{\cite{acharya2024quantum}}    & Tab. S7, subgrid (0,8)  \\
                                      & 5 & $2.52 \times 10^{-3}$ &                                 &                                               & Fig. S18, 72Q device  \\
                                      & 7 & $1.43 \times 10^{-3}$ &                                 &                                               & Main text  \\ \hline
        \multirow{2}{*}{Surface code} & 3 & $5.80 \times 10^{-3}$ & \multirow{2}{*}{Harmony, RL prior} & \multirow{2}{*}{\cite{eickbusch2025demonstratingdynamicsurfacecodes}} & Tab. F9, avg. over $d=3$ \\
                                      & 5 & $2.70 \times 10^{-3}$ &                                    &                                                                       & Tab. F9  \\ \hline
        \multirow{3}{*}{Surface code} & 3 & $4.00 \times 10^{-3}$ & \multirow{3}{*}{Neural network} & \multirow{3}{*}{This work} & Tab.~\ref{tab:decoder_error_rates}, subgrid (8,8)  \\ 
                                      & 5 & $1.58 \times 10^{-3}$ &                                 &                            & Tab.~\ref{tab:decoder_error_rates}, subgrid (0,4)  \\ 
                                      & 7 & $0.77 \times 10^{-3}$ &                                 &                            & Tab.~\ref{tab:decoder_error_rates} \\ \hline
        \multirow{2}{*}{Color code}   & 3 & $1.72 \times 10^{-2}$ & \multirow{2}{*}{Neural network} & \multirow{2}{*}{\cite{lacroix2025scaling}} & Extracted from $\varepsilon_{5}, \Lambda_{3/5}$  \\
                                      & 5 & $1.10 \times 10^{-2}$ &                                 &                                            & Main text \\ \hline
        Color code                    & 5 & $0.82 \times 10^{-2}$ & Tesseract, freq. prior          & This work                  & Main text \\ \hline
    \end{tabular}
    \caption{{\bf Summary of the best published LERs,} obtained with the most accurate decoders in the respective references. All LER values are averaged over $X$ and $Z$ bases. These results are shown as empty symbols in Fig.~\ref{fig_sm:google_history}.}
    \label{tab:best_qec_results}
\end{table}

\begin{table}
    \centering
    \renewcommand{\arraystretch}{1.3} 
    \setlength{\tabcolsep}{8pt}       
    \begin{tabular}{|c|c|c|c|}
        \hline
        \textbf{QEC code} & \textbf{Distance} & \textbf{LER} & \textbf{Reference} \\ 
        \hline \hline 
        \multirow{3}{*}{Surface code} & 3 & $6.63 \times 10^{-3}$ & \multirow{3}{*}{\cite{acharya2024quantum}}  \\
         & 5 & $3.48 \times 10^{-3}$ &  \\
         & 7 & $2.27 \times 10^{-3}$ &  \\ \hline
        \multirow{2}{*}{Surface code} & 3 & $6.68 \times 10^{-3}$ & \multirow{2}{*}{\cite{eickbusch2025demonstratingdynamicsurfacecodes}}  \\ 
         & 5 & $3.20 \times 10^{-3}$ &  \\ \hline
        \multirow{3}{*}{Surface code} & 3 & $4.95 \times 10^{-3}$ & \multirow{3}{*}{This work}  \\
         & 5 & $2.46 \times 10^{-3}$ &  \\
         & 7 & $1.42 \times 10^{-3}$ &  \\ \hline
        Color code & 5 & $1.27 \times 10^{-2}$ & \cite{lacroix2025scaling} \\ \hline
        Color code & 5 & $0.92 \times 10^{-2}$ & This work \\ \hline
    \end{tabular}
    \caption{{\bf Summary of LERs with consistent decoding.} Surface code is decoded with Sparse Blossom, and color code with Tesseract, both using SI1000 prior. The experimental datasets are the same as in Table~\ref{tab:best_qec_results} and are publicly available as indicated in Table~\ref{tab:published_results}. These results are shown as solid symbols connected with lines in Fig.~\ref{fig_sm:google_history}.} 
    \label{tab:results_with_equiv_decoding}
\end{table}

\subsection{AlphaQubit2}
\label{sec:aq2}

We used the AlphaQubit2 (AQ2) neural network decoder \cite{senior2025scalable} for the surface code only, see Section~\ref{sec:distance7}. The results of AQ2 decoding are shown in Fig.~\ref{fig_sm:d7_training_run}(e,f) and are summarized in Table~\ref{tab:decoder_error_rates}.

The AQ2 model was employed with several augmentations and a significantly simpler, less compute-intensive training protocol. While previous works \cite{bausch2023learning,acharya2024quantum,lacroix2025scaling,senior2025scalable} required training separate models for each subgrid and logical basis, we have now trained a single model to decode in both $X$ and $Z$ bases across all distances ($d=3, 5, 7$), including all respective subgrid locations (nine locations for $d=3$, four for $d=5$, and one for $d=7$). 

Previously, we utilized a three-stage training protocol: (1) pretraining on synthetic data sampled from the SI1000 error model in Stim; (2) further training on synthetic data from a DEM fitted to experimental data (specialized by distance, basis, and subgrid); and (3) fine-tuning on a dedicated experimental dataset. By contrast, in this work the increased volume of available fine-tuning data and the move to a unified model allowed us to omit the intermediate stage and focus only on (1) and (3). The fine-tuning dataset is similar to that in Fig.~\ref{fig_sm:d7_training_run}(f) and was acquired in an experiment preceding the final data acquisition. A separate, similar dataset was used for validation, while the final data in Fig.~\ref{fig_sm:d7_training_run}(f) was decoded once after completing the fine-tuning and validation.

We augmented the AQ2 model from Ref.~\cite{senior2025scalable} in several ways:
\begin{itemize}
    \item In addition to predicting the final logical observable, we predict a label for each cycle of the experiment. This is equivalent to the noiseless intermediates predicted by the standard AQ2 but implemented via an incremental logical observable in the DEM, similar to \cite{zhang2026learning}. As with previous AQ1 and AQ2 implementations, we predict $d$ parallel (and redundant) logical observables for each cycle in a distance-$d$ code.
    \item To enable using a single model across all QEC circuit instances, we provide the logical basis, each stabilizer basis, physical stabilizer coordinates, and code distance as model inputs. The final model layer features two outputs -- one for each logical basis. 
    \item We further augmented the input by embedding the average error detection rate for each stabilizer at every cycle, averaged across the validation set. This allows the model to better adjust to physical noise levels. The physical coordinate embeddings allow the model to also adapt to noise variations across different subgrid locations.
\end{itemize}

Final predictions are formed by an ensemble of four models fine-tuned from two independently pretrained ones. The model utilizes the temporal compression as in Ref.~\cite{senior2025scalable} (combining five consecutive cycles before processing). We pretrained for up to 4 billion shots, sampling uniformly across the three code distances and across noise levels ranging from $0.25$ to $3$ times the device noise level determined by the average EDR of the training data for each distance, subgrid, cycle count, and basis. For fine-tuning and testing, we selected the best-performing models based on the validation metrics of the pretraining and fine-tuning stages, respectively.

    \section{Learning algorithm}
\label{sec:rl_algo}

In this section, we provide a pedagogical derivation of the RL algorithm. To highlight the broad applicability of our framework, we maintain an abstract discussion using generalized terminology such as {\it system}, {\it controller}, {\it observations}, etc. We then contextualize these concepts by mapping them directly to our QEC optimization problem, illustrating just one of many potential applications of this RL framework. The pseudocode of the learning algorithm in the QEC context is shown in Algorithm~\ref{alg:rl_qec}.

Many aspects of the derivation follow established techniques from the RL literature. In particular, our approach is heavily inspired by the parameter-exploring policy gradient \cite{sehnke2010parameter} and is enhanced by ideas from proximal policy optimization \cite{schulman2017proximal}. We additionally incorporate techniques such as entropy regularization \cite{haarnoja2018soft}, replay buffer \cite{mnih2015human}, and gradient masking. The derivation builds upon Appendix~G of Ref.~\cite{sivak2024optimization}, albeit using different conceptual interpretation. 

\subsection{Problem setup}

Consider the system controller parametrized with vector $\p\in \mathbb{R}^P$, where $P$ is the total number of control parameters, which in practice can be several thousands, see Tab.~\ref{tab:policy_size}. We refer to $\p$ as the {\it control policy}. A learning algorithm explores the control space by sampling candidate policies from some distribution $\pi(\p|\boldsymbol{\theta})$, where $\boldsymbol{\theta}$ denotes the distribution parameters. We refer to it as the {\it policy distribution}, but it is also known as the search distribution in related literature on evolutionary algorithms \cite{wierstra2014natural}. 
It is worth contrasting this setup with textbook RL frameworks \cite{sutton2018reinforcement} that typically model sequential decision-making over discrete time steps within an episode, and where policy specifies a probability distribution over the actions taken at those time steps. The insight of Ref.~\cite{sehnke2010parameter} is that it is possible to reformulate the mathematical framework of RL by transferring the exploration from the action space directly to the policy space, an approach particularly suitable for our QEC problem. 

When the system controller behaves according to a policy $\p$, the observations $\o\in\mathbb{R}^O$ received from the system contain implicit information about the quality of this policy. We denote the  distribution function of observations as $\varepsilon(\o|\p): \mathbb{R}^P \to \mathbb{R}^O$, and note that this distribution is conditionally independent of $\boldsymbol{\theta}$ given $\p$. In RL, this function is known as the {\it environment} of the learning agent. It is often assumed that the dynamics of the environment is so complex that it is infeasible to model, in which case we only have query access to $\varepsilon$. In this sense, the RL approach can be considered model-free, although some minimal modeling assumptions will enter the framework later on.

Finally, given the observations $\o$ from the environment, the reward for the policy $\p$ is assigned using the reward function $r(\o): \mathbb{R}^O\to \mathbb{R}$. This function quantifies how the observed system behavior differs from the desired behavior, allowing to rank the policy candidates. For example, in the present work the observations are the error detection rates of all detectors in the QEC circuit, and the reward is minus the total error detection rate, $r(\o)=-\one^T\o$. 

The objective of the learning agent is to find the policy $\p_{\rm opt}$ that maximizes the reward. In RL, this is achieved by optimizing the performance measure $J(\boldsymbol{\theta})$, defined as the expected reward when sampling the policy candidates from the distribution $\pi(\p|\boldsymbol{\theta})$:
\begin{align}
    J(\boldsymbol{\theta}) = \underset{\pi(\p|{\boldsymbol{\theta}})}{\mathbb{E}}\;\underset{\varepsilon(\o|\p)}{\mathbb{E}}[r(\o)].
    \label{performance_measure}
\end{align}

During the learning process, the agent adjusts the learnable parameters $\boldsymbol{\theta}$ to optimize this measure. In a stationary environment, where  $\varepsilon(\o|\p)$ does not change over time, the policy distribution evolves towards a deterministic distribution $\pi_{\rm opt}(\p) = \delta(\p - \p_{\rm opt})$ that always returns $\p_{\rm opt}$ (assuming that optimal policy is unique and that the parametrization admits this $\delta$-distribution in some limit).
In the policy-gradient RL \cite{sutton2018reinforcement}, to find the optimal policy starting from an arbitrary initial distribution $\pi(\p|\boldsymbol{\theta})$, we estimate the gradient $\nabla_{\boldsymbol{\theta}} J(\boldsymbol{\theta})$ with Monte Carlo rollouts and iteratively update $\boldsymbol{\theta}$ until convergence. 

In this work, we use a simple factorized Gaussian distribution:
\begin{align}
    \pi(\p|\boldsymbol{\theta}) = \frac{\exp\left(-\frac{1}{2}(\p-\boldsymbol{\mu})^T\boldsymbol{\Sigma}^{-1}(\p-\boldsymbol{\mu})\right)}{\sqrt{(2\pi)^P\det(\boldsymbol{\Sigma}) }},  \label{gaussian_policy}
\end{align}
with mean $\boldsymbol{\mu}$ and diagonal covariance matrix $\boldsymbol{\Sigma}={\rm diag}[\boldsymbol{\sigma}^2]$. The learnable parameters of this policy distribution are $\boldsymbol{\theta} = (\boldsymbol{\mu}, \boldsymbol{\sigma})$. Note that our derivation is directly compatible with a more general class of distributions that satisfy the factorization property $\pi(\p|\boldsymbol{\theta})=\prod_{i=1}^P \pi_i(p_i|\boldsymbol{\theta})$. 

Furthermore, we assume that the reward function $r(\o)$ has a special structure: it is a sum of $O$ local terms, where each term only depends on the corresponding observation component. Such reward function can be conveniently represented as $r(\o)=\one^T\r$ using the {\it reward vector} $\r\in\mathbb{R}^O$. For example, in the present work we used $\r=-\o$, meaning that the RL agent's goal is to minimize the total rate of error detection events.

The final assumption we use is that each observation component $o_j$ and, by extension, the reward component $r_j$ depends only on a restricted subset of the policy parameters $\{p_i| i\in{\cal S}_j\}$, where the subset ${\cal S}_j$ enumerates the relevant parameters. For efficient convergence, it is crucial that this set remains bounded and does not scale with the system size, i.e. $\forall j: |{\cal S}_j|=O(1)$ as $P\to\infty$. 
In the QEC context, this assumption means that the number of control parameters per gate and the number of gates per detecting region does not increase as the system size grows, which is well aligned with the principles of fault-tolerant QEC circuit constructions. In some cases, e.g. in quantum systems with significant gate crosstalk \cite{wesdorp2026mitigating}, this assumption is violated, and the algorithm's convergence rate can depend on $P$, in contrast to crosstalk-free result in Fig.~5(b-c).

\subsection{Gradient estimator}

Next, we describe the construction of the gradient estimator. We require this estimator to be unbiased, i.e. converge to the true gradient $\nabla_{\boldsymbol{\theta}} J(\boldsymbol{\theta})$ in the limit of infinite samples from the policy distribution. In addition, to facilitate efficient learning in practice, especially when environment queries are expensive, an estimator with minimal variance is desired \cite{mohamed2020monte}.  

We begin the derivation by transforming the gradient of the performance measure in Eq.~\eqref{performance_measure}:
\begin{align}
    \nabla_{\boldsymbol{\theta}} J(\boldsymbol{\theta}) 
    = \nabla_{\boldsymbol{\theta}}\, \underset{\pi(\p|{\boldsymbol{\theta}})}{\mathbb{E}}\;\underset{\varepsilon(\o|\p)}{\mathbb{E}}[r(\o)] 
    = \nabla_{\boldsymbol{\theta}} \int \underset{\varepsilon(\o|\p)}{\mathbb{E}}[r(\o)]\,  \pi(\p|\boldsymbol{\theta}) d\p
    = \int \underset{\varepsilon(\o|\p)}{\mathbb{E}}[r(\o)]\, \nabla_{\boldsymbol{\theta}} \pi(\p|\boldsymbol{\theta}) d\p \label{grad_integral}.
\end{align}

Keeping the estimator unbiased, we can subtract an arbitrary {\it baseline} $b$ from $r(\o)$ in Eq.~\eqref{grad_integral}, as long as this baseline does not depend on $\p$. This is the case because $\int b\,  \nabla_{\boldsymbol{\theta}} \pi(\p|\boldsymbol{\theta}) d\p=b\,  \nabla_{\boldsymbol{\theta}} \int \pi(\p|\boldsymbol{\theta}) d\p = b  \nabla_{\boldsymbol{\theta}} 1 = 0$. Note that the baseline $b$ can be arbitrary, even a random variable, and it can depend on the policy parameters $\boldsymbol{\theta}$. Introducing the baseline is a common variance reduction technique known as control variate \cite{mohamed2020monte}.
Intuitively, the optimal baseline approximately corresponds to the expected reward under the current policy. We will discuss how to choose $b$ in more detail shortly, but for now we impose that $b$ has a similar decomposition to $r(\o)$, i.e. it is a sum of $O$ terms corresponding to each observation component, $b=\one^T\b$, where $\b\in\mathbb{R}^O$.

Next, we define the {\it advantage} function
\begin{align}
    \boldsymbol{\alpha}(\o) = \r(\o) - \b,
    \label{advantage}
\end{align}
such that policy candidates with rewards larger than the baseline have a positive advantage, while those with reward smaller than baseline have negative advantage.
The gradient in Eq.~\eqref{grad_integral} can be transformed using these new objects:
\begin{align}
    \nabla_{\boldsymbol{\theta}} J(\boldsymbol{\theta}) 
    & = \int \underset{\varepsilon(\o|\p)}{\mathbb{E}}[\one^T \boldsymbol{\alpha}(\o)]\, \nabla_{\boldsymbol{\theta}} \pi(\p|\boldsymbol{\theta}) d\p.
\end{align}

We shall represent it in the form of an expectation value over the samples produced from some distribution. Instead of sampling $\p\sim \pi(\p|\boldsymbol{\theta})$, we take samples from a potentially different {\it collection distribution} $\pi(\p|\tilde{\boldsymbol{\theta}})$ with variables $\tilde{\boldsymbol{\theta}}$. The use of importance sampling here results in a PPO type gradient estimator \cite{schulman2017proximal}, while if the samples were taken from the actual $\pi(\p|\boldsymbol{\theta})$ distribution then we would obtain a \textsc{reinforce}-type estimator \cite{williams1992simple}, also known as the score-function estimator \cite{mohamed2020monte}. 
Now the gradient becomes:
\begin{align}
    \nabla_{\boldsymbol{\theta}} J(\boldsymbol{\theta}) 
    & = \int \underset{\varepsilon(\o|\p)}{\mathbb{E}}[\one^T \boldsymbol{\alpha}(\o)]\, \frac{\nabla_{\boldsymbol{\theta}} \pi(\p|\boldsymbol{\theta})}{\pi(\p|\boldsymbol{\tilde{\theta}})} \pi(\p|\boldsymbol{\tilde{\theta}}) d\p
    \nonumber\\
    &=\nabla_{\boldsymbol{\theta}}\underset{\pi(\p|\boldsymbol{\tilde{\theta}})}{\mathbb{E}}\,\underset{\varepsilon(\o|\p)}{\mathbb{E}}\Big[\one^T \boldsymbol{\alpha}(\o)\, \frac{\pi(\p|\boldsymbol{\theta})}{\pi(\p|\boldsymbol{\tilde{\theta}})} \Big]
    \nonumber\\
    &=\nabla_{\boldsymbol{\theta}}\underset{\pi(\p|\boldsymbol{\tilde{\theta}})}{\mathbb{E}}\,\underset{\varepsilon(\o|\p)}{\mathbb{E}}\Big[\sum_{j=1}^O \alpha_j(o_j)\prod_{i=1}^P\frac{\pi_i(p_i|\boldsymbol{\theta})}{\pi_i(p_i|\boldsymbol{\tilde{\theta}})} \Big],
\end{align}
where in the last line we made use of our two structural assumptions: that the advantage function $\boldsymbol{\alpha}$ is decomposable as a sum of $O$ local terms, and that the policy $\pi$ factorizes over the space of $P$ control parameters.  

Note that the quantity inside the expectation operator is the product of an advantage function computed for a given policy and observation sample $(\p, \o)$, and the importance ratio for this policy sample under two different distributions. Recall an additional assumption in our framework that observations and, by extension, the rewards and advantages have only local dependence on the policy parameters, i.e. the observation component $o_j$ depends on parameters $\{p_i |i\in {\cal S}_j\}$. This means that importance ratio can be simplified to include only relevant policy components, which allows to drastically reduce its variance while maintaining the same expectation value: 
\begin{align}
    \nabla_{\boldsymbol{\theta}} J(\boldsymbol{\theta})  &=\nabla_{\boldsymbol{\theta}}\mathbb{E}\Big[\sum_{j=1}^O \alpha_j(o_j)\prod_{i\in{\cal S}_j}\frac{\pi_i(p_i|\boldsymbol{\theta})}{\pi_i(p_i|\boldsymbol{\tilde{\theta}})} \Big],
\end{align}
where we adopted a shorthand notation for the full expectation operator $\mathbb{E}[\cdot] \equiv \underset{\pi(\p|{\boldsymbol{\tilde{\theta}}})}{\mathbb{E}}\;\;\underset{\varepsilon(\o|\p)}{\mathbb{E}}[\cdot]$.

To simplify this expression further, it is convenient to introduce the importance ratio vector $\boldsymbol{\chi}_{\boldsymbol{\theta}\boldsymbol{\tilde{\theta}}}(\p)\in\mathbb{R}^O$, such that $\chi_i=\pi_i(p_i|\boldsymbol{\theta})/\pi_i(p_i|\boldsymbol{\tilde{\theta}})$, and the masking matrix $\M\in\mathbb{R}^{O\times P}$ defined such that $M_{ji}$ is $1$ if $i \in {\cal S}_j$ and $0$ otherwise. 
This matrix endows the learning algorithm with minimal knowledge about the environment  function $\varepsilon(\o|\p)$: only its sparse structure of dependencies but not their actual functional form. {\it This is a single modeling assumption of our RL framework that requires some prior knowledge about the system.} If this knowledge is lacking, one can always replace $0$'s with $1$'s as a conservative measure: any such replacement would still result in an unbiased gradient estimator albeit of higher variance (even the limiting case $\M={1}^{O\times P}$, equivalent to global optimization). We emphasize that harnessing the sparsity of $\M$ is critical for scalability of the learning algorithm.

The gradient can now be concisely written as
\begin{align}
    \nabla_{\boldsymbol{\theta}} J(\boldsymbol{\theta})  &=\nabla_{\boldsymbol{\theta}}\mathbb{E}\Big[\boldsymbol{\alpha}(\o)^T \exp(\M\times \ln\boldsymbol{\chi}_{\boldsymbol{\theta}\boldsymbol{\tilde{\theta}}}(\p)) \Big],
    \label{nice_gradient}
\end{align}

In practice, we obtain its empirical estimator by collecting a batch of Monte Carlo rollouts $(\p^{(k)}, \o^{(k)}, \r^{(k)})$ for $k\in 1,...,B$, where $B$ is the batch size, and replacing the exact expectation value in Eq.~\eqref{nice_gradient} with the sample mean. In principle, the collection policy $\pi(\p|\boldsymbol{\tilde{\theta}})$ used to acquire the batch could be arbitrary. However, to achieve stable and efficient learning it is important for this policy to not differ significantly from the current policy $\pi(\p|\boldsymbol{\theta})$, otherwise the importance ratios $\chi_i$ will strongly deviate from 1, which will increase the estimator variance and can result in training instabilities \cite{schulman2017proximal}. Hence, we simply use the ``old'' policy with parameters $\boldsymbol{\tilde{\theta}}$ from the most recent learning epoch. In some cases, samples from several epochs can be accumulated into the replay buffer, although we observed mixed success of this approach in the non-stationary setting. 

\subsection{Loss functions}

In modern machine learning frameworks the gradients are computed via automatic differentiation. We can leverage this infrastructure by defining the following loss functions.

From Eq.~\eqref{nice_gradient} we extract the main {\it policy loss function}:
\begin{align}
    L_{\rm policy} = -\hat{\mathbb{E}}\Big[\boldsymbol{\alpha}(\o)^T \exp(\M\times \ln\boldsymbol{\chi}_{\boldsymbol{\theta}\boldsymbol{\theta'}}(\p)) \Big],
    \label{eq:policy_loss}
\end{align}
where the importance ratio vector $\boldsymbol{\chi}$ is clipped element-wise near 1, as proposed in Ref.~\cite{schulman2017proximal} for improved stability.

In addition to the parameters $\boldsymbol{\theta}$ of the policy distribution, we optimize the baseline $\b$ using the {\it baseline loss function}:
\begin{align}
    L_{\rm baseline} = \hat{\mathbb{E}}\Big[\|\boldsymbol{\alpha}(\o)\|^2\Big],
    \label{eq:value_loss}
\end{align}
which, according to the definition in Eq.~\eqref{advantage}, simply encourages the baseline to smoothly track the moving average of the reward vector. 

Finally, to facilitate exploration, which is especially important in non-stationary environments to prevent premature convergence, we include the {\it entropy regularization loss function} 
\cite{haarnoja2018soft}:
\begin{align}
  L_{\rm entropy} = - \hat{\mathbb{E}}\Big[ H(\pi(\p|\boldsymbol{\theta})) \Big]
    \label{eq:entropy_loss}
\end{align}
that favors increasing the entropy of the policy distribution, defined as
\begin{align}
  H(\pi(\p|\boldsymbol{\theta})) = -\int \pi(\p|\boldsymbol{\theta}) \ln \pi(\p|\boldsymbol{\theta}) d\p.
\end{align}

The total loss is therefore:
\begin{align}
    L_{\rm total} = L_{\rm policy} + L_{\rm baseline} + L_{\rm entropy}.
    \label{total_loss}
\end{align}

For the Gaussian policy distribution in Eq.~\eqref{gaussian_policy}, there are a total of $2P+O$ learnable parameters that are directly optimized with gradient descent on this loss function: $P$ components of $\boldsymbol{\mu}$, $P$ components of $\boldsymbol{\sigma}$, and $O$ components of the baseline $\b$, making the algorithm extremely lightweight. Moreover, with the Gaussian policy the gradients of all losses in Eq.~\eqref{total_loss} have simple analytic expressions and do not require any back-propagation. Due to these favorable properties, our algorithm is amenable to efficient on-FPGA implementations which could drastically improve calibration data throughput \cite{marciniak2026millisecond}. In our actual experiment, the agent was implemented on a host CPU.

Finally, note that the behavior of the learning algorithm is regulated using {\it hyperparameters}, such as the learning rate, gradient clipping magnitude, relative weights for different loss components in Eq.~\eqref{total_loss}, etc. We performed shallow manual optimization of hyperparameters using a combination of real and simulated RL runs, and expect that further performance gains could be achieved with automated hyperparameter tuning frameworks.

\begin{figure}
\addtocounter{figure}{-1} 
\renewcommand{\figurename}{Algorithm}
\renewcommand{\thefigure}{1} 
\noindent\rule{\linewidth}{0.8pt}
\caption{RL calibration of quantum control parameters for QEC}
\label{alg:rl_qec}
\vspace{2pt}
\noindent\rule{\linewidth}{0.8pt}
\vspace{-8pt} 
\begin{algorithmic}[1]
\State \textbf{Input:} Masking matrix $\boldsymbol{M}$ defining detector-parameter dependencies
\State \textbf{Initialize:} Policy parameters $\boldsymbol{\theta} = (\boldsymbol{\mu}, \boldsymbol{\sigma})$ and baseline vector $\boldsymbol{b}$
\For{epoch $t = 1, 2, \dots, T$}
    \State Set collection policy parameters $\boldsymbol{\tilde{\theta}} \leftarrow \boldsymbol{\theta}$ (treated as constant)
    \State Sample a batch of $B$ candidate policies $\{\boldsymbol{p}^{(k)}\}_{k=1}^B$ from $\pi(\boldsymbol{p}|\boldsymbol{\tilde{\theta}})$
    \For{each candidate $k \in \{1, \dots, B\}$}
        \State Apply physical control parameters $\boldsymbol{p}^{(k)}$ to the classical controller
        \State Sample observations $\boldsymbol{o}^{(k)}$ from the environment and compute rewards $\boldsymbol{r}^{(k)}$
    \EndFor
    \State Compute the total loss $L_{total}$ for the data batch $\{(\boldsymbol{p}^{(k)}, \boldsymbol{o}^{(k)}, \boldsymbol{r}^{(k)})\}_{k=1}^B$ \Comment{Ref. Eqs.~\eqref{eq:policy_loss}, \eqref{eq:value_loss}, \eqref{eq:entropy_loss}, and \eqref{total_loss}}
    \State Update $\boldsymbol{\theta}$ and $\boldsymbol{b}$ via gradient descent: $\boldsymbol{\theta}, \boldsymbol{b} \leftarrow \text{OptimizerStep}(\nabla_{\boldsymbol{\theta}, \boldsymbol{b}} L_{total})$
\EndFor
\end{algorithmic}
\vspace{-8pt} 
\noindent\rule{\linewidth}{0.8pt}
\end{figure}

    \clearpage
    \bibliographystyle{apsrevlongbib}
    \bibliography{bib.bib}